\documentclass[a4paper, 12pt]{article}
\usepackage{amstext}
\usepackage{amsmath}
\usepackage{amsthm}
\usepackage{mathtools}
\usepackage{physics}
\usepackage{mathrsfs}
\usepackage{array}
\usepackage{titling}
\usepackage{enumitem}
\usepackage[dvipsnames]{xcolor}
\usepackage[calc, en-US]{datetime2}

\usepackage[labelsep=period]{caption}
\usepackage{subcaption}

\usepackage[page]{appendix}

\usepackage{enumitem}
\usepackage{algorithm}
\usepackage{algpseudocode}
\usepackage{graphicx}
\graphicspath{{graphs/}}

\usepackage[margin=1.25in]{geometry}
\usepackage[onehalfspacing]{setspace}

\usepackage[scaled=1,helvratio=1]{newtxtext}
\usepackage[varg,vvarbb]{newtxmath}
\usepackage[varqu,varl]{inconsolata}
\usepackage[type1,scaled=1]{cabin}

\usepackage{anyfontsize}
\usepackage{bm}

\DeclareMathOperator{\ind}{\mathbf{1}}
\DeclareMathOperator*{\argmax}{arg\,max}
\newcommand{\me}{\mathrm{e}}
\newcommand{\R}{\mathbb{R}}

\DeclareMathOperator{\E}{\mathrm{E}}
\DeclareMathOperator{\Prob}{\mathrm{P}}
\DeclareMathOperator{\Lgen}{\mathrm{L}}

\theoremstyle{plain}

\newtheorem{lemma}{Lemma}
\newtheorem{proposition}{Proposition}
\newtheorem{corollary}{Corollary}
\newtheorem{assumption}{Assumption}

\theoremstyle{definition}
\newtheorem{definition}{Definition}

\theoremstyle{remark}
\newtheorem*{remark}{Remark}

\usepackage[
backend=biber,
sorting=nyt,
citestyle=authoryear,
bibstyle=authortitle,
doi=false,
isbn=false,
url=false,
eprint=false
]{biblatex}
\usepackage{hyperref}
\hypersetup{
    colorlinks,
    linkcolor = Blue,
    citecolor = Blue,
    urlcolor = Black
}

\pretitle{\begin{center}\LARGE\bfseries}
\posttitle{\par\end{center}\vskip 1.5em

\par
}

\preauthor{\begin{center}
    \large \vskip 0.5em
    }
\postauthor{\par\end{center}}

\predate{\begin{center}\small\normalfont\itshape}
\postdate{\par\end{center}}

\title{Strategic Exploration for Innovation}
\author{
    Shangen Li\thanks{
        \protect Zurich Center for Market Design, University of Zurich (e-mail address: \url{work@shangen.li}).
        \protect I am grateful to
Marlon Azinovic,
Heski Bar-Isaac,
Steven Callander,
Jiahua Che,
Martin Cripps,
Christian Ewerhart,
Marina Halac,
David H\'{e}mous,
Johannes H\"{o}rner,
Bruno Jullien,
Vijay Krishna,
Nicolas Lambert,
Alessandro Lizzeri,
Margaret Meyer,
Nick Netzer,
Sven Rady,
Anne-Katrin Roesler,
Johannes Schneider,
Jakub Steiner,
Bruno Strulovici,
Yu Fu Wong,
Leeat Yariv,
Haoyuan Zeng,
and seminar participants
at the University of Zurich,
CEIBS,
and Swiss Theory Day 2021
for helpful comments.
I am especially indebted to
Marek Pycia, Armin Schmutzler, 
and Aleksei Smirnov
for advice and encouragement.
This paper is based on the second chapter of my Ph.D.\ dissertation at the University of Zurich.

    }
}

\date{\DTMdate{2023-12-12}}

\begin{document}

\begin{titlepage}
    \maketitle

    \begin{abstract}
    This paper introduces a framework to study innovation in a strategic setting,
    in which innovators allocate their resources between exploration and exploitation in continuous time.
    Exploration creates public knowledge,
    while exploitation delivers private benefits.
    Through the analysis of a class of Markov equilibria,
    we demonstrate that knowledge spillovers accelerate knowledge creation and expedite its availability,
    thereby encouraging innovators to increase exploration.
    The prospect of the ensuing superior long-term innovations further motivates exploration,
    giving rise to a positive feedback loop.
    This novel feedback loop can substantially mitigate the free-riding problem arising from knowledge spillovers.
\end{abstract}

    
\end{titlepage}

\section{Introduction}

Without innovation, we would still be living in caves and hunting with stones.
Although people celebrate disruptive inventions like penicillin and the Internet,
innovations are often the result of small improvements through incremental experimentation.
This approach can be observed in various fields
and leads to significant advancements and innovations.
In agriculture,
breeding and selection of crops
have led to new varieties with improved yields and disease resistance.
In architecture,
iconic structures that balance form and function have emerged
through refinement of styles over time.
And the gradual improvement of digital electronics
has resulted in powerful, portable, and user-friendly devices.

The pursuit of innovation through experimentation
typically entails opportunity costs and uncertainties,
resulting in ubiquitous trade-offs
between creating value through innovation (exploration)
and capturing value through established operations (exploitation).
As innovation is not guaranteed,
innovators must carefully weigh the potential benefits of exploring new alternatives
against the costs of diverting resources from their proven operations.
As a result,
incentives for experimentation can be undermined by free-riding
when its outcomes are publicly observable,
or when technologies can be reverse-engineered.
Such information and knowledge spillovers
lead to learning inefficiencies
and stifle technological progress.
Given the prevalence of free-riding opportunities,
one might ask why anyone would pursue exploration.
What are the trade-offs that individuals and organizations face
when developing technologies with strategic considerations in mind?
How does this strategic effect impact technological progress in the long run?
We explore these questions
in a model intended to capture the dynamics of collective exploration
in a strategic setting,
often seen in contexts such as
research joint ventures, open-source software development, and academic collaborations.

This paper analyzes a game of strategic exploration
in which a finite number of forward-looking players
jointly search for innovative technologies.
Given a set of publicly available technologies,
at each point in time,
each player decides
how to allocate a perfectly divisible unit of resource
between exploration, 
which expands the set of feasible technologies
at a rate proportional to the resource allocated,
and exploitation,
which yields a flow payoff
from the adoption of one of the feasible technologies.
The qualities of technologies,
which determine the flow payoff from exploitation,
are represented by an (initially unknown) realized path of Brownian motion,
referred to as the \emph{technological landscape}.
Not all technologies are readily available,
and only the qualities associated with the explored technologies are known.
The outcomes of exploration are transparent,
and the technologies developed are treated as public goods.
Consequently, 
perfect information and knowledge spillovers occur among players,
allowing some to benefit from the positive externalities generated by the experimentation of others,
thereby giving rise to abundant opportunities for free-riding.

One of the major contributions of this paper 
is to extend the classic game of strategic experimentation
to a setting with an unboundedly expandable set of arms
while allowing for a certain degree of correlation among them.
When modeling technologies as arms,
correlation between arms
and discoveries of new arms
are key features
of technology development and innovation.
However,
most of the literature on strategic experimentation
assumes a fixed set of independent arms.
Indeed,
the analysis in the previously studied models
would be substantially complicated by correlation.
By contrast,
our proposed model demonstrates that
under a form of correlation structure
that is simple yet appropriate in the context of technology development,
such complexities can be circumvented by simply
focusing on incremental experimentation.
This approach offers an intuitive way
to capture the richness of the dynamic and path-dependent nature
of technology development.

We first consider the efficient benchmark
in which the players work jointly to maximize the average payoff.
The solution to this cooperative problem
takes a simple cutoff form:
All players allocate their resource exclusively to exploration
if the quality difference between
the best available technology
and the latest outcome of the experiment
is below a time-invariant cutoff,
and exclusively to exploitation otherwise.
Players in a larger team are more ambitious,
manifested by their persistence in exploration
even after experiencing a prolonged series of setbacks.
As team size grows,
in the limit,
exploration persists even when the quality difference becomes arbitrarily large,
indefinitely broadening the set of long-run feasible technologies.

In the strategic problem,
we restrict attention to Markov perfect equilibria (MPE)
with the difference between the qualities of
the best available technology
and the latest technology under development
as the state variable.
Despite the considerable disparities
between our setting and the two-armed bandit models in the literature on strategic experimentation,
all MPE in our model exhibit a similar \emph{encouragement effect}:
The future information and knowledge spillovers from other players encourage at least one of them
to continue exploration
at states where a single agent would already have given up.
The players thus act more ambitiously in the hope that
successful outcomes will bring the other players back to exploration in the future,
promoting innovation and sharing its burden.
In addition,
exactly because of such an incentive,
any player who never explores
would strictly prefer the deviation
that resumes the exploration process
as a volunteer
at the state where all exploration stops.
Therefore,
no player always free-rides in equilibrium
even though
exploration per se never produces any payoffs.

We further show that
there is no MPE in which all players use simple cutoff strategies
as in the cooperative solution.
This result suggests that in any symmetric equilibrium,
each player chooses
an interior allocation of their resource at some states.
We establish the existence and uniqueness of the symmetric equilibrium,
and provide closed-form representations
for the equilibrium strategies and the associated payoff functions.
Within the region of interior allocation,
the players gradually reduce the resource allocated to exploration
as the latest outcomes deteriorate.
Nevertheless,
exploration never fully stops
in the symmetric equilibrium.
As the latest outcomes keep deteriorating,
exploration would slow down so severely that
technologies never progress
to the point at which all players
prefer to allocate the resource exclusively to exploitation.
This observation strongly suggests that
asymmetric equilibria with
the rate of exploration bounded away from zero
in the region of interior allocation
could improve welfare
over the symmetric equilibrium.
To investigate such a possibility,
we construct a class of asymmetric MPE
in which the players take turns performing exploration
at unpromising states,
so that each player
achieves a higher payoff than in the symmetric equilibrium.
It turns out that these asymmetric MPE are the best MPE
in two-player games
in terms of average payoffs.

Above all,
we identify a novel \emph{innovation effect}
that is essential for understanding the incentives for technology development in dynamic and strategic environments.
To explain this effect,
we introduce the concept of \emph{returns to cooperation}
as follows.
We say a technological landscape yields \emph{decreasing} returns to cooperation
if players' payoff functions in the cooperative solution remain bounded as the number of players increases to infinity,
reminiscent of the existing bandit models with a fixed set of arms.
By contrast,
exploration cooperatively over landscapes
with nondecreasing returns to cooperation
gives rise to unbounded payoffs in the limit.
This asymptotic feature sets our model apart from the existing ones,
as it demonstrates that introducing innovation into multi-armed bandits models
allows us to identify a novel incentive for exploration: the prospect of technological advancement.
We then examine the interplay between this innovation effect and the encouragement effect
through the comparative statics of the unique symmetric equilibrium
with respect to the number of players.
Our analysis shows that the encouragement effect and the innovation effect reinforce each other.
Moreover, 
the innovation effect boosts
the encouragement effect to overcome the free-rider effect as the team gets large,
if and only if the players are sufficiently patient
and the underlying landscape yields \emph{increasing} returns to cooperation.
In such a case,
even though the rate of exploration is suboptimal,
the set of long-run feasible technologies expands indefinitely as the team size grows,
resembling the cooperative solution.
The prevalence of the encouragement effect in our model
stands in marked contrast to the existing literature on strategic experimentation,
in which the free-rider effect always prevails
due to the lack of innovation possibilities.

\subsection{Related Literature}
\label{sec:lit}

This paper combines two distinct strands of literature on learning and experimentation.
The first strand studies experimentation in rich and complex environments,
drawing on the seminal work by
\textcite{Callander:2011}.
He proposed modeling
the correlation between technologies by Brownian path
and studies experimentation conducted by a sequence of myopic agents.
In his model,
experiment outcomes closer to zero are preferable to the agents.
In a similar setting,
\textcite{CallanderMatouschek:2019} 
consider agents with insatiable preferences
and examine the impact of risk aversion 
on their search performance.
\textcite{GarfagniniStrulovici:2016} extend Callander's model
to a setting with overlapping generations,
in which short-lived yet forward-looking players
search on a Brownian path for technologies with higher qualities.
They mainly focus on
the search patterns and the long-run dynamics,
and establish the stagnation of search
and the emergence of a technological standard in finite time.
The qualities of technologies contribute exponentially to the payoffs
in our model instead of linearly as in theirs.
As a result, stagnation can be avoided in our model even 
under a negative drift of the Brownian path.
All these models focus on non-strategic environments
and preclude long-lived forward-looking agents,
as
their discrete-time frameworks create unexplored gaps between explored technologies,
posing challenges for further analysis.
To overcome this difficulty,
we forgo the fine details of the learning dynamics
for analytical tractability
by imposing continuity on the experimentation process.
This simplification can be interpreted as
the qualities of the neighboring technology
being revealed during experimentation,
so that no unexplored territory remains between the explored technologies.
Moreover,
we impose a hard constraint on the scope of exploration
to capture the scenario that
technologies far ahead of their time are infeasible to be explored today.
These abstractions
allow us to derive 
explicit expressions for the equilibrium payoffs and strategies,
perform comparative statics analysis,
and construct asymmetric equilibria.

The second strand of literature,
often referred to as strategic experimentation,
originated from the Brownian model introduced by \textcite{BoltonHarris:1999},
and further enriched by the exponential model in
\textcite{KellerRadyCripps:2005}
and the Poisson model in \textcite{KellerRady:2010,KellerRady:2015}.
In all of these models,
players face identical two-armed bandit machines,
which consist of
a risky arm with an unknown quality
and a safe arm with a known quality.
At each point in time,
each player decides
how to split one unit of a perfectly divisible resource
between these arms,
so that learning occurs gradually
by observing other players' actions and outcomes.
These models differ in the assumptions
on the probability distribution of the flow payoffs
that each type of arm generates.
By contrast,
players in our model face
a continuum of correlated arms.
Local learning---learning the quality of a particular arm---occurs instantaneously.
However,
as the set of arms is unbounded,
global learning---learning the qualities of all arms---occurs gradually.

The encouragement effect was first identified by \textcite{BoltonHarris:1999}
in the symmetric equilibrium in their Brownian model.
This effect is then established by \textcite{KellerRady:2010}
for all MPE in the Poisson model with inconclusive good news,
and by \textcite{KellerRady:2015}
for the symmetric MPE in the Poisson model with bad news.
Due to the absence of technological advancements,
the encouragement effects in all these papers
are not strong enough to dominate the free-rider effect.\footnote{
    \textcite{BoltonHarris:1999} demonstrate the prevalence of the free-rider effect in their comparative statics analysis.
    They show that in the symmetric equilibrium of their model,
    the individual resource allocated to experimentation at beliefs below the myopic cutoff converge to zero as the number of players increases. 
    The same feature can be observed in the symmetric MPE of the Poisson model in
    \textcite{KellerRady:2010} and \textcite{KellerRady:2015},
    despite the absence of a formal comparative statics analysis in those papers.
}
We demonstrate not only the presence of the encouragement effect in all MPE of our model,
but also perform comparative statics for the unique symmetric MPE
to further investigate the strength of the encouragement effect.
In contrast to the encouragement effect in their models,
we find that the prospect of innovation and technological advancements
enables the encouragement effect to overcome the free-rider effect,
and provide the conditions for this to occur.

More broadly, this paper contributes to the literature
on dynamic public-good games.
\textcite{AdmatiPerry:1991},
\textcite{MarxMatthews:2000},
\textcite{Yildirim:2006},
and \textcite{Georgiadis:2014}
study voluntary contributions to a joint project in dynamic settings.
While the public good in these papers is the progress toward the completion of a project,
the public good in our model
is the knowledge---the feasible technologies---built over time,
which can be exploited once developed.

From a modeling perspective,
continuous exploration on a Brownian sample path
is independently studied by
\textcite{Wong:2022} and \textcite{UrgunYariv:2023}
in non-strategic environments.
To the best of our knowledge,
the only other study of collective exploration on a Brownian path
in a strategic setting is an independent work by
\textcite{CetemenUrgunYariv:2023}.
They focus on the exit patterns during a search process conducted jointly by heterogeneous players.
In their model,
exploitation is only possible after an irreversible exit chosen endogenously by each player.
The players in our model, however, are not faced with stopping problems
and thus are free to choose
between exploration and exploitation, or even both simultaneously, at all times.

\section{The Exploration Game}
\label{sec:model}
Time \(t\in[0,\infty)\) is continuous, and the discount rate is \(r>0\).
There are \(N\geq 1\) players,
each endowed with
one unit of perfectly divisible resource per unit of time.
Each player has to independently allocate
her resource between
exploration,
which expands the feasible technology domain,
and exploitation,
which allows her to adopt one of the explored technologies.
The feasible technology domain,
which is common to all players and
contains all the explored technologies at time \(t\),
is modeled as an interval \([0, X_t]\) with \(X_0=0\).
If a player allocates the fraction \(k_t\in[0,1]\) to exploration
over an interval of time \([t, t+\dd{t})\),
the boundary \(X_t\)
is pushed to the right
by an amount of \(k_t \dd{t}\).
With the fraction \(1-k_t\) allocated to exploitation,
by adopting technology $x_t\in[0, X_t]$,
the player receives a deterministic flow payoff \((1-k_t)\exp(W(x_t))\dd{t}\),
where \(W(x_t)\) denotes the quality of the adopted technology.

The technological landscape 
\(W:\R_{+}\to \R\),
which maps technologies to their qualities,
is common to all players.
Nevertheless, only the qualities of the feasible technologies in \([0, X_t]\)
are known to each player at time \(t\).
The status quo technology \(X_0 = 0\)
has a quality \(W(0) = s_0\),
whereas the qualities of the initially unexplored technologies
on \(\R_{++}\)
are specified by a realized path of Brownian motion
in \(C(\R_{+}, \R)\)
starting at \(w_0 = W(0+)\leq s_0\),
with drift \(\mu\in\R\)
and volatility \(\sigma > 0\).\footnote{
    In other words,
    the mapping \(W\) is a realized Brownian path on \(\R_+\),
    except possibly for a discontinuity at the origin with \(W(0) \geq W(0+)\).
}

At the outset of the game,
all players know the parameters of the landscape
\(\mu, \sigma, w_0\) and \(s_0\),
but not the realized Brownian path.
Therefore,
the process of exploration described above
captures the dynamics of
\emph{research}---experimenting with unknown technologies---and \emph{development}---expanding the set of feasible technologies.
Collective exploration as such thus features
both \emph{information} and \emph{knowledge} spillovers.\footnote{
    Dynamic games featuring pure information spillover include
    bandits-based games such as in \textcite{BoltonHarris:1999,KellerRadyCripps:2005},
    in which all technologies are feasible at the outset, but their qualities are uncertain and to be learned.
    By contrast,
    dynamic games featuring pure knowledge spillover can be thought of as
    games of public goods provision
    such as in \textcite{AdmatiPerry:1991, MarxMatthews:2000, Yildirim:2006,Georgiadis:2014},
    in which players contribute to a joint project (e.g., developing a technology in the public domain) with its value known in advance.
    Neither of these two types of model fully captures the progressive and uncertain nature of the underlying process of technology development and innovation.

}

Given a player's actions \(\{(k_t, x_t)\}_{t\geq 0}\),
with \(k_t\in[0,1]\) and \(x_t\in[0,X_t]\)
measurable with respect to the information available at time \(t\),
her total expected discounted payoff,
expressed in per-period units, is
\[
    {\E}\left[
        \int_0^\infty  r \me^{-rt} (1-k_t) \me^{W(x_t)} \dd{t}
    \right].
\]

Note that whenever a player chooses exploitation,
she always adopts one of the best feasible technologies
\(x_t \in \argmax_{x\in[0,X_t]}W(x)\)
to maximize her total expected discounted payoff.
Therefore, we can focus on such an exploitation strategy without loss
and rewrite the above total payoff as
\[
    {\E}\left[
        \int_0^\infty r \me^{-rt} (1-k_t) \me^{S_t} \dd{t} 
    \right],
\]
where \(S_t = \max_{x\in[0,X_t]}W(x)\).

\subsection{Reformulated Game}
\label{sec:model-reformulated}

The environment above can be equivalently reformulated as follows.
Players have prior beliefs represented by a filtered probability space \((\Omega, \mathscr{F}, (\mathscr{F}_t), {\Prob})\),
where \(\Omega = C(\R_{+}, \R)\)
is the space of Brownian paths,
\({\Prob}\) is the law of standard Brownian motion \(B = \{B_t\}_{t\geq 0}\),
and \(\mathscr{F}_t\) is the canonical filtration of \(B\).
Each player chooses her strategy
from the space of admissible control processes \(\mathcal{A}\),
which consists of all processes \(\{k_{t}\}_{t\geq 0}\)
adapted to the filtration \((\mathscr{F}_t)_{t\geq 0}\)
with \(k_{t}\in[0,1]\).
The public history of technology development
is represented by 
the process \(\{W(X_t)\}_{t > 0}\),
which is the original Brownian motion under the time change controlled by the players' strategies.
This process satisfies the stochastic differential equation
\[
    \dd{W(X_t)} = \mu K_t\dd{t} + \sigma\sqrt{K_t} \dd{B_t},\quad W(0+) = w_0,
\]
where \(K_t = \sum_{1\leq n \leq N} k_{n,t}\)
measures how much of the overall resource is allocated to exploration,
and will be referred to as the \emph{intensity of exploration} at time \(t\).

Given a strategy profile \(\bm{k} = \{(k_{1,t},\ldots,k_{N,t})\}_{t\geq 0}\),
player \(n\)'s total expected discounted payoff can be written as
\[
    {\E}\left[
        \int_0^\infty r \me^{-rt} (1-k_{n,t}) \me^{S_t} \dd{t}
    \right],
\]
where
\[
    S_t = \max_{0\leq\tau\leq t}W(X_\tau)
\]
denotes the quality of the best feasible technology at time \(t\).

In addition,
we use the term ``gap'',
denoted by \(A_t \coloneqq S_t - W(X_t)\) for \(t > 0\),
whereas by \(A_0 \coloneqq s_0 - w_0\) for \(t = 0\),
to refer to the quality difference between
the best feasible technology
and
the latest technology under development.
Henceforth,
we shall use \(a\) and \(s\) when referring to the state variables
as opposed to the stochastic processes
\(\{A_t\}_{t\geq 0}\) and \(\{S_t\}_{t\geq 0}\)
(i.e., if \(A_t = a\),
then ``the game is in state \(a\) at time \(t\)'').

A Markov strategy \(k_n:\R_+\times\R\to [0,1]\)
with \((a,s)\) as the state variable
specifies the action player \(n\) takes at time \(t\) to be \(k_n(A_t, S_t)\).
A Markov strategy is called \emph{\(s\)-invariant}
if it depends on \((a,s)\)
only through \(a\).
Thus an \(s\)-invariant Markov strategy \(k_n:\R_+\to[0,1]\)
takes the gap \(a\) as the state variable.
Finally,
an \(s\)-invariant Markov strategy \(k_n\) is a \emph{cutoff strategy}
if there is a cutoff \(\bar{a}\geq 0\) such that
\(k_n(a) = 1\) for all \(a \in[0,\bar{a})\) and \(k_n(a) = 0\) otherwise.

Given an \(s\)-invariant Markov strategy profile \(\bm{k}\),
the homogeneity of the payoff functions
enables us to write player \(n\)'s associated payoff at state \((a,s)\)
as
\(v_n(\, a,s \mid \bm{k} \,) = \me^s v_n(\, a, 0 \mid \bm{k} \,)\).\footnote{
    See Lemma~\ref{lem:homogeneity} in Appendix~\ref{sec:proofs-ppt-payoff-functions}.
}
It is thus convenient to define
\(u_n(\, a \mid \bm{k} \,) \coloneqq v_n(\, a, 0 \mid \bm{k} \,)\),
which equals player \(n\)'s payoff at state \((a, s)\)
normalized by \(\me^s\), the opportunity cost of exploration.
We refer to \(u_n:\R_{+}\to\R\) 
as player \(n\)'s \emph{normalized payoff function},
or simply as \emph{payoff function}
when it is clear from the context.

Further, 
let \(\theta\coloneqq 2\mu/\sigma^2\)
and \(\rho\coloneqq\sigma^2/(2r)\).
It can be shown that
each player's expected discounted payoff remains finite
as long as the following condition is satisfied.
\begin{assumption} \label{asm:main-assumption}
    \(N\rho(1+\theta) < 1\).
\end{assumption}
To ensure well-defined payoff functions and deviation,
we impose Assumption~\ref{asm:main-assumption} for the remainder of the paper unless otherwise stated.

\section{Joint Maximization of Average Payoffs}
\label{sec:cooperative-problem}
Suppose that \(N\geq 1\) players
work cooperatively to maximize the \emph{average} expected payoff.
Denote by \(\mathcal{A}_N\)
the space of all adapted processes \(\{K_t\}_{t\geq 0}\)
with \(K_t\in[0,N]\).
Formally,
we are looking for
the value function
\[
    v(a,s) \coloneqq \sup_{K\in\mathcal{A}_N} v(\, a,s \mid K \,),
\]
where
\[
    v(\, a,s \mid K \,) \coloneqq {\E}_{as}\left[
            \int_0^\infty r \me^{-rt} (1-K_t/N) \me^{S_t} \dd{t}
    \right]
\]
denotes the average payoff function
associated with the control process \(K = \{K_t\}_{t\geq 0}\),
and an optimal control \(K^*\in\mathcal{A}_N\)
such that \(v(a,s) = v(\, a,s \mid K^* \,)\).
The structure of the problem
allows us to focus on Markov strategies
\(K:\R_+\times\R\to [0,N]\)
with \((a,s)\) as the state variable,\footnote{
    For now,
    the existence of a Markovian optimal strategy is still a conjecture.
    It will be verified later in Proposition~\ref{prop:cooperative-solution}.
}
so that the intensity of exploration at time \(t\)
is specified by \(K_t = K(A_t, S_t)\).

According to the dynamic programming principle, we have
\begin{equation*}
    v(a,s)
    = \max_{K\in[0,N]}\left\{
        r \left(1 - K/N\right) \me^s\dd{t} + {\E}_{as}\left[\me^{-r\dd{t}} v(a+\dd{a}, s+\dd{s})\right]
        \right\}.
\end{equation*}
First, note that \(S_t\) can only change when \(A_t = 0\),
and thus \(\dd{s} = 0\) for all positive gaps.
Hence, for each \(a > 0\) at which \(\pdv*[2]{v}{a}\) is continuous,
the value function \(v\) satisfies the Hamilton-Jacobi-Bellman (HJB) equation
\begin{equation}
    \label{eq:hjb-coop-xs}
    v(a,s) = \max_{K\in[0,N]}\left\{
        \left(1 - \frac{K}{N}\right) \me^s + K\rho\left(
        {\pdv[2]{v(a,s)}{a}}
        -\theta {\pdv{v(a,s)}{a}}
        \right)
    \right\}.
\end{equation}

Assume, as will be verified,
that the optimal strategy is \(s\)-invariant.
Then by the homogeneity of the value function
we can replace \(v(a,s)\) with \(\me^s u(a)\),
and divide both sides of equation~\eqref{eq:hjb-coop-xs} by the opportunity cost \(\me^s\),
to obtain the normalized HJB equation
\begin{equation}
    \label{eq:hjb-coop}
    u(a) = 1 + \max_{K\in[0,N]} K\{\beta(a, u) - 1/N\},
\end{equation}
where \(\beta(a, u) \coloneqq \rho(u''(a)-\theta u'(a))\)
is the ratio of the expected benefit of exploration
\(\rho(\pdv*[2]{v}{a} - \theta\pdv*{v}{a})\)
to its opportunity cost \(\me^s\).
It is then straightforward to see
that the optimal action takes the following ``bang-bang'' form.
If the shared opportunity cost of exploration, \(1/N\),
exceeds the full expected benefit,
the optimal choice is \(K(a) = 0\) (all agents choose exploitation exclusively),
which gives \(u(a) = 1\).
Otherwise, \(K(a) = N\) is optimal (all agents choose exploration exclusively),
and \(u\) satisfies the second-order ordinary differential equation (henceforth ODE),
\begin{equation}
    \label{eq:ode-coop}
    \beta(a, u) = u(a)/N.
\end{equation}

The optimal strategy could presumably depend on both \(a\) and \(s\)
and hence might not be \(s\)-invariant.
Indeed, both
the benefit and the opportunity cost of exploration
increase as innovation occurs.
Nevertheless,
due to our specific form of flow payoff
where the qualities of technologies contribute exponentially to the payoffs,
the increased benefit of exploration
exactly offsets the increased opportunity cost.
As a result,
the incentives for exploration at any fixed gap
do not depend on the highest-known quality,
which leads to an \(s\)-invariant optimal strategy.
This conjecture is confirmed by the following proposition.

\begin{proposition}[Cooperative Solution]
    \label{prop:cooperative-solution}
    Suppose Assumption~\ref{asm:main-assumption} holds.
    In the \(N\)-agent cooperative problem,
    there is a cutoff \(a^*_N > 0\) given by
    \begin{equation*}
        a^*_N = \frac{1}{\gamma_2 - \gamma_1}\left(
            \ln\left(1+\frac{1}{\gamma_2}\right)
            -
            \ln\left(1+\frac{1}{\gamma_1}\right)
        \right)
    \end{equation*}
    with \(\gamma_1 < \gamma_2\)
    being the roots of
    \(
        \gamma(\gamma - \theta) = 1/(N\rho),
    \)
    such that
    it is optimal for all players to choose exploitation exclusively
    when the gap
    is above the cutoff \(a^*_N\) 
    and
    it is optimal for all players to choose exploration exclusively
    when the gap is below the cutoff \(a^*_N\).

    The associated payoff at state \((a,s)\)
    can be written as \(V^*_N(a,s) = \me^s U^*_N(a)\),
    where the normalized payoff function \(U^*_N:\R_{+}\to\R\)
    is given by
    \begin{equation*}
        U^*_N(a) = \frac{1}{\gamma_2 - \gamma_1}
                \left(
                    \gamma_2 \me^{-\gamma_1 (a^*_N - a)}
                    -\gamma_1 \me^{-\gamma_2(a^*_N - a)}
                \right)
    \end{equation*}
    when \(a \in [0, a^*_N)\),
    and by \(U^*_N(a) = 1\) otherwise.

    If Assumption~\ref{asm:main-assumption} is violated,
    then \(U^*_N(a) = +\infty\) for all \(a\geq 0\).
\end{proposition}

The cooperative solution is pinned down by 
the standard smooth pasting condition \(u'(a^*_N) = 0\),
and the \emph{normal reflection} condition \((\pdv*{v}{a} + \pdv*{v}{s})(0+,s) = 0\),
which takes the form of \(u(0) + u'(0+) = 0\)
for \(s\)-invariant strategies.\footnote{
    The normal reflection condition
    is not an optimality condition.
    It ensures that the infinitesimal change of the payoff at a zero gap
    has a zero \(\dd{S}\) term,
    which is necessary for the continuation value process to be a martingale.
    See \textcite{PeskirShiryaev:2006} for an introduction
    to the normal reflection condition in the context of optimal stopping problems,
    and the proof of Lemma~\ref{lem:payoff-function-ppt-x}
    for more details.
}

Because of the lack of information on the qualities of technologies,
the players might stop too early, thereby ultimately adopting a technology that is inferior to the one they would have pursued had its quality been known in advance,
or might stop too late,
wasting too many resources for marginal improvement.
The cooperative solution
optimally balances these trade-offs
between early and late stopping
and therefore determines the \emph{efficient} strategies.

\subsection{Cooperative Solution: Comparative Statics}

To facilitate the discussion on the comparative statics
as well as our analysis in later sections,
we now introduce the concept of \emph{returns to cooperation} associated with a technological landscape as follows.
\begin{definition}
    We say (the prior distribution of) a technological landscape \(W\) yields
    \begin{itemize}
        \item 
        \emph{decreasing returns to cooperation (DRC)} 
        if \(\theta < -1\),
        \item \emph{constant returns to cooperation (CRC)}
        if \(\theta = -1\),
        \item 
        and \emph{increasing returns to cooperation (IRC)}
        if \(\theta > -1\).
    \end{itemize}
\end{definition}

Note that Assumption~\ref{asm:main-assumption}
is always satisfied when \(W\) yields DRC or CRC.

\begin{corollary}
    \label{cor:cs-coop}
    The cooperative cutoff \(a^*_N\) is strictly increasing in \(N\).
    For each \(\theta\) and \(\rho\),
    \begin{itemize}
        \item if \(W\) yields DRC,
        then \(\lim U^*_N < +\infty\) as \(N\to +\infty\);
        \item if \(W\) yields CRC,
        then \(U^*_N\to +\infty\) as \(N\to +\infty\);
        \item if \(W\) yields IRC,
        then \(U^*_N\to +\infty\) as \(N\to 1/(\rho(1+\theta))\).\footnote{
            Here we allow \(N\) to be non-integral values for convenience.
            Also note that when \(W\) yields IRC,
            Assumption~\ref{asm:main-assumption}
            is violated for \(N\geq 1/(\rho(1+\theta))\),
            in which case \(U^*_N = +\infty\).
        }
    \end{itemize}
    In all cases \(a^*_N\to +\infty\).
\end{corollary}

A larger stopping cutoff \(a^*_N\)
represents greater ambition among the players.
Because the players work cooperatively,
as the team size increases, 
extra resources brought by additional players
enable a higher rate of exploration.
Consequently, for each fixed level of ambition,
resources are wasted for a shorter period of time
before the exploration fully stops,
which motivates the players to become more ambitious.
This leads to the emergence of more advanced technologies in the long run
and further drives the players to act more ambitiously,
and so forth.

Corollary~\ref{cor:cs-coop} underlines
a key component absent in the literature on bandits-based games:
innovation.
The set of feasible technologies (arms) is predetermined and fixed in
the Brownian model in \textcite{BoltonHarris:1999} and the Poisson model in \textcite{KellerRady:2010}.
Even though additional players expedite learning,
the discounted payoff stream derived from technology adoption is bounded
due to a bounded technology space.
As a result,
each player's payoff eventually maxes out as the team size grows.
In our model,
technology adoption still offers a bounded payoff stream.
However,
as more players join exploration,
they become more ambitious
and thus more advanced technologies emerge in the long run.
Whether this prospect of superior innovations
qualitatively alters the asymptotic property of players' payoffs
depends on the returns to cooperation associated with the technological landscape.
As characterized in Corollary~\ref{cor:cs-coop},
collective exploration over
a DRC landscape 
entails diminishing marginal welfare improvement with respect to team size,
and thus resembles the bandits-based games with a fixed set of arms,
whereas a CRC or IRC landscape unleashes the power of innovation as team gets large.\footnote{
    As shown in Corollary~\ref{cor:cs-coop},
    \(U_N^*\to +\infty\) under both CRC and IRC landscapes.
    What differentiates these two cases is that
    \(U_N^* < +\infty\) for any finite \(N\) under a CRC landscape,
    whereas \(U_N^*\to+\infty\) as \(N\) approaches some finite number under an IRC landscape.
}
This \emph{innovation effect} has important implications in the upcoming analysis of the strategic problem.

\section{The Strategic Problem}

From now on,
we assume that there are \(N > 1\) players acting noncooperatively.
We study equilibria in the class of
\(s\)-invariant Markov strategies,
which are the Markov strategies with the gap as the state variable
and will hereafter be referred to as \emph{Markov strategies}.
In this section,
we provide characterizations of the best responses
and the associated payoff functions,
which establish
useful properties of the equilibria
for further analysis.

\subsection{Best Responses and Equilibria}

We denote by \(\mathcal{K}\) the set of Markov strategies
that are right-continuous and piecewise Lipschitz-continuous,
and denote by \(\mathcal{A}\)
the space of admissible control processes
as in Section~\ref{sec:model-reformulated}.\footnote{
    Piecewise Lipschitz-continuity means that
    \(\R_{+}\) can be partitioned
    into a finite number of intervals
    such that the strategy is Lipschitz-continuous on each of them.
    This requirement rules out the infinite-switching strategies considered in Section 6.2 of \textcite{KellerRadyCripps:2005}.
}
A strategy \(k_n^*\in\mathcal{K}\)
for player \(n\) is a best response against
her opponents' strategies
\(\bm{k}_{\neg n} = (k_1,\ldots,k_{n-1},k_{n+1},\ldots,k_N)\in\mathcal{K}^{N-1}\)
if
\[
    v_n(\, a,s \mid k_n^*, \bm{k}_{\neg n} \,)
    = \sup_{k_n\in\mathcal{A}} v_n(\, a,s \mid k_n, \bm{k}_{\neg n} \,)
\]
at each state \((a,s)\in\R_+\times\R\).
This definition turns out to be equivalent to
\[
    u_n(\, a \mid k_n^*, \bm{k}_{\neg n} \,)
    = \sup_{k_n\in\mathcal{K}} u_n(\, a \mid k_n, \bm{k}_{\neg n} \,)
\]
for each gap \(a\geq 0\),
with the normalized payoff function
\(u_n(\, a \mid \bm{k} \,)\)
defined as in Section~\ref{sec:model-reformulated}.
A Markov perfect equilibrium is a profile of Markov strategies
that are mutually best responses.

Denote the intensity of exploration
carried out by player \(n\)'s opponents
by \(K_{\neg n}(a) = \sum_{l\neq n}k_l(a)\),
and the benefit-cost ratio of exploration
by \(\beta(a, u_n)\)
as in Section~\ref{sec:cooperative-problem}.
The following lemma 
characterizes all MPE in the exploration game.

\begin{lemma}[Equilibrium Characterization]
    \label{lem:char-mpe}

A strategy profile \(\bm{k} = (k^*_1,\ldots,k^*_N)\in\mathcal{K}^N\)
is a Markov perfect equilibrium
with 
\(u_n:\R_{+}\to \R\)
being the corresponding payoff function of player \(n\) for each \(n \in\{1,\ldots, N\}\),
if and only if
for each \(n\),
function \(u_n\)  
\begin{enumerate}[label=(\roman*)]
    \item is continuous on \(\R_+\) and once continuously differentiable on \(\R_{++}\);\label{item:eq-char-1}
    \item is piecewise twice continuously differentiable on \(\R_{++}\);\footnote{
        This condition means that there is partition of \(\R_{++}\) into a finite number of intervals such that
        \(u_n''\) is continuous on the interior of each of them.
        }
    \label{item:eq-char-2}
    \item satisfies the normal reflection condition 
    \begin{equation}
        \label{eq:normal-reflection}
        u_n(0) + u_n'(0+) = 0;
    \end{equation}
    \label{item:eq-char-3}
    \item satisfies,
    at each continuity point of \(u_n''\),
    the HJB equation
    \begin{equation}\label{eq:hjb}
        u_n(a) = 1 + K_{\neg n}(a) \beta(a, u_n) + \max_{k_n\in[0, 1]} k_n\{\beta(a, u_n) - 1\},
    \end{equation}
    \label{item:eq-char-4}
\end{enumerate}
with \(k^*_n(a)\) achieving the maximum on
the right-hand side, i.e.,
\[
    k^*_n(a) \in 
    \argmax_{k_n\in[0,1]}
    k_n\{\beta(a, u_n) - 1\}.
\]
\end{lemma}

These conditions are standard in optimal control problems.
Condition~\ref{item:eq-char-4},
and the smooth pasting condition
implicitly stated in Condition~\ref{item:eq-char-1},
are the optimality conditions.
The remaining conditions pertain to payoff functions in general.

In any MPE,
Lemma~\ref{lem:char-mpe} provides
the following characterization of best responses.
If \(\beta(a, u_n) < 1\),
then \(k_n^*(a) = 0\) is optimal and
\(u_n(a) = 1 + K_{\neg n}(a) \beta(a, u_n) < 1 + K_{\neg n}(a)\).
If \(\beta(a, u_n) = 1\),
then the optimal \(k_n^*(a)\) takes arbitrary values in \([0,1]\) and
\(u_n(a) = 1 + K_{\neg n}(a)\).
Finally, if \(\beta(a, u_n) > 1\),
then \(k_n^*(a) = 1\) is optimal and
\(u_n(a) = (1 + K_{\neg n}(a)) \beta(a, u_n) > 1 + K_{\neg n}(a)\).
In short,
player \(n\)'s best response to
a given intensity of exploration \(K_{\neg n}\)
by the others
depends on whether \(u_n\) is 
greater than, equal to, or less than \(1 + K_{\neg n}\).

On the intervals where each \(k_n\) is continuous,
HJB equation~\eqref{eq:hjb} gives rise to the ODE
\begin{equation}
    \label{eq:feynman-kac-x}
    u_n(a) = 1 - k_n(a) + K(a)\beta(a,u_n).
\end{equation}
In particular,
on the intervals where each \(k_n\) is constant,
the ODE above admits the explicit solution
\begin{equation}
    \label{eq:sol-full-intensity}
    U(a) = 1 - k_n + C_1 \me^{\gamma_1 a} + C_2 \me^{\gamma_2 a},
\end{equation}
where \(\gamma_1\) and \(\gamma_2\)
are the roots of the equation \(\gamma(\gamma - \theta) = 1/(K\rho)\),
and \(C_1, C_2\) are constants to be determined.

Lastly,
on the intervals where an interior allocation is chosen by player \(n\),
the ODE from the indifference condition \(\beta(a, u_n) = 1\) has the
general solution
\begin{equation}
    \label{eq:sol-int-alloc}
    U(a) = \begin{cases}
        C_1 + C_2 a +  a^2/(2\rho), &\text{ if } \theta = 0, \\
        C_1 + C_2 \me^{\theta a} - a/(\rho\theta), &\text{ if } \theta\neq 0,
    \end{cases}
\end{equation}
where \(C_1, C_2\) are constants to be determined.

\subsection{Properties of MPE}
\label{sec:mpe-ppt}

First, note that in any MPE,
the average payoff can never exceed 
the \(N\)-player cooperative payoff \(U_N^*\),
and no individual payoff can fall below
the single-agent payoff \(U_1^*\).
The upper bound follows directly from the fact that
the cooperative solution maximizes the average payoff.
The lower bound \(U^*_1\) is guaranteed
by playing the single-agent optimal strategy,
as the players can only benefit
from the exploration efforts of others.

Second, all Markov perfect equilibria are inefficient.
Along the efficient exploration path,
the benefit of exploration tends to \(1/N\) of its opportunity cost
as the gap \(A_t\)
approaches the efficient stopping threshold.
A self-interested player thus
has an incentive to deviate to exploitation
whenever the benefit of exploration drops below its full opportunity cost.

Note also that in any MPE,
the set of states at which
the intensity of exploration is positive
must be an interval \([0,\bar{a})\)
with \(a^*_1 \leq \bar{a}\leq a^*_N\).
The bounds on the stopping threshold 
follow directly from the bounds on the average payoffs
and imply that
the technologies emerging in the long run
in any MPE
cannot outperform the cooperative solution.

Moreover,
the intensity of exploration must be bounded away from zero
on any compact subset of \([0,\bar{a})\).
If this were not the case,
there would exist some gap \(a < \bar{a}\)
such that the process
\(\{W(X_t)\}_{t > 0}\) starting from \(w_0 = s_0-a\)
would never reach the best-known quality \(s_0\)
because of diminishing intensity,
and therefore
allocating a positive fraction of the resource to exploration at gap \(a\)
is clearly not optimal for any player.

In the two-armed bandit models reviewed in Section~\ref{sec:lit},
the players always use the risky arm
at beliefs higher than some myopic cutoff,
above which the expected short-run payoff from the risky arm
exceeds the deterministic flow payoff from the safe arm.
Because our model lacks such a myopic cutoff,
it might seem reasonable to conjecture that
some player \(n\),
with her payoff function \(u_n\) bounded from above by \(1+K_{\neg n}\),
never explores,
and thus free-rides the technologies developed by the other players.
Such a conjecture is refuted by the following proposition.

\begin{proposition}[No Player Always Free-rides]
    \label{prop:everyone-explores}
    In any Markov perfect equilibrium,
    no player allocates her resource exclusively to exploitation for all gaps.
\end{proposition}

The intuition behind this result is that
in equilibrium,
the cost and benefit of exploration
must be equalized at the stopping threshold
for each player,
whereas any player who never explores
would find this benefit outweighs the cost,
manifested by a kink in her payoff function at the stopping threshold.
She would then have a strict incentive to resume exploration
immediately after the other players give up,
hoping to reduce the gap
to bring the exploration process back alive.
Therefore, in equilibrium,
every player must perform exploration at some states,
respecting the smooth pasting condition (Condition~\ref{item:eq-char-1} in Lemma~\ref{lem:char-mpe}).

This result shows that
each player strictly benefits from the presence of the other players
in equilibrium through information and knowledge spillovers.
As a result,
the future exploration efforts of the others encourage some of the players
to explore at gaps larger than their single-agent cutoffs.
Such an \emph{encouragement effect} is exhibited in all MPE of the exploration game.

\begin{proposition}[Encouragement Effect]
    \label{prop:encouragement-effct}
    In any Markov perfect equilibrium,
    at least one player explores at gaps
    above the single-agent cutoff \(a^*_1\).
\end{proposition}

With the same intuition as the encouragement effect,
our last general result on Markov perfect equilibria
concerns the nonexistence of equilibria where all players use cutoff strategies.
\begin{proposition}[No MPE in Cutoff Strategies]
    \label{prop:no-cutoff-strategies}
    In any Markov perfect equilibrium, at least one player uses a strategy
    that is not of the cutoff type.
\end{proposition}

Next, we look into Markov perfect equilibria in greater depth.

\section{Symmetric Equilibrium}

Our characterization of best responses
and the nonexistence of MPE in cutoff strategies
suggest that in any symmetric equilibrium,
the players choose an interior allocation at some states.
At these states of interior allocation,
the benefit of exploration
must be equal to the opportunity cost,
and therefore the common payoff function solves
the ODE \(\beta(a, u) = 1\).
As a consequence of equation~\eqref{eq:feynman-kac-x},
the payoff of each player at the states
of interior allocation in the symmetric equilibrium
must also satisfy
\(u = 1 + K_{\neg n} \leq N\).
Therefore,
whenever the common payoff exceeds \(N\),
each player allocates her resource exclusively to exploration,
and the payoff function satisfies the same ODE~\eqref{eq:ode-coop}
as in the cooperative solution.
However,
it is worth pointing out that
for some configurations of the parameters,
because of the strength of free-riding incentives among the players,
the common payoff could be below \(N\) for all gaps,
and accordingly,
the resource constraint of each player
is not necessarily binding
even when the gap is zero,
in marked contrast to the cooperative solution.
Lastly, the common payoff satisfies \(u = 1\) at the gaps
for which the resource is exclusively allocated to exploitation.

The solutions to the corresponding ODEs provided
in equations~\eqref{eq:sol-full-intensity} and~\eqref{eq:sol-int-alloc},
together with the normal reflection condition~\eqref{eq:normal-reflection}
and the smoothness requirement on the equilibrium payoff functions,
uniquely pin down the strategies and the associated payoff functions
in the symmetric equilibrium,
which can be expressed in closed form as follows.

\begin{proposition}[Symmetric Equilibrium]
    \label{prop:symmetric-mpe}
    The \(N\)-player exploration game has a unique
    symmetric Markov perfect equilibrium
    with the gap as the state variable.
    There exists a stopping threshold \(\tilde{a}_N \in (a^*_1, a^*_N)\)
    and a full-intensity threshold \(a^\dagger_N\geq 0\) such that
    the fraction \(k^\dagger_N(a)\) of the resource
    that each player allocates to exploration at gap \(a\)
    is given by
    \begin{equation}
        k^\dagger_N(a) = \begin{cases}
            0, & \text{ on } [\tilde{a}_N,+\infty), \\
            \frac{1}{(N-1)\rho}\int_0^{\tilde{a}_N-a}\phi_\theta(z)\dd{z} \in(0,1), & \text{ on } [a^\dagger_N, \tilde{a}_N), \label{eq:symmetric-MPE}\\
            1, & \text{ on } [0,a^\dagger_N) \text{ if } a^\dagger_N > 0,
        \end{cases}
    \end{equation}
    with \(\phi_\theta(z) \coloneqq (1-\me^{-\theta z})/\theta\).\footnote{
        We define \(\phi_0(z) \coloneqq \lim_{\theta\to 0}\phi_\theta(z) = z\).
    }

    The corresponding payoff function  is the unique function 
    \(U^\dagger_N:\R_{+}\to[1,+\infty)\)
    of class \(C^1\) with the following properties:
    \(U^\dagger_N(a) = 1\) on \([\tilde{a}_N,+\infty)\);
    \(U^\dagger_N(a) = 1+(N-1)k^\dagger_N(a)\in(1, N)\) 
    and solves the ODE \(\beta(a, u) = 1\)
    on \((a^\dagger_N,\tilde{a}_N)\);
    if \(a^\dagger_N > 0\),
    then \(U^\dagger_N(a) > N\)
    and solves the ODE \(\beta(a, u) = u/N\) on \((0, a^\dagger_N)\).

\end{proposition}
The closed-form expressions for the common payoff function \(U^\dagger_N\)
and the thresholds \(a^\dagger_N\) and \(\tilde{a}_N\)
are provided in Appendix~\ref{sec:sym-mpe-expression}.

\begin{figure}[htb]
    \centering
    \includegraphics[width=\textwidth]{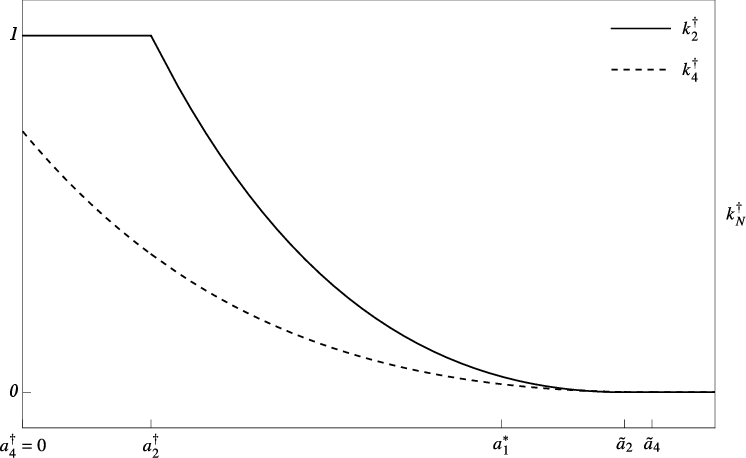}
    \caption{
    The symmetric equilibrium
    with binding resource constraints
    in a two-player game,
    and with non-binding constraints
    in a four-player game
    (\(\rho = 2\), \(\theta = -1\)).}
    \label{fig:kplotsym}
\end{figure}

As we have already pointed out,
depending on the parameters,
it is possible that \(a^\dagger_N = 0\),
in which case \(k^\dagger_N(a) < 1\) for all \(a > 0\),
and hence will be referred to as the \emph{non-binding case}.
The opposite case, where \(a^\dagger_N > 0\), will be referred to as the \emph{binding case}.
Figure~\ref{fig:kplotsym} illustrates
the symmetric equilibrium for these two cases.

\begin{figure}[htb]
    \centering
    \includegraphics[width=\textwidth]{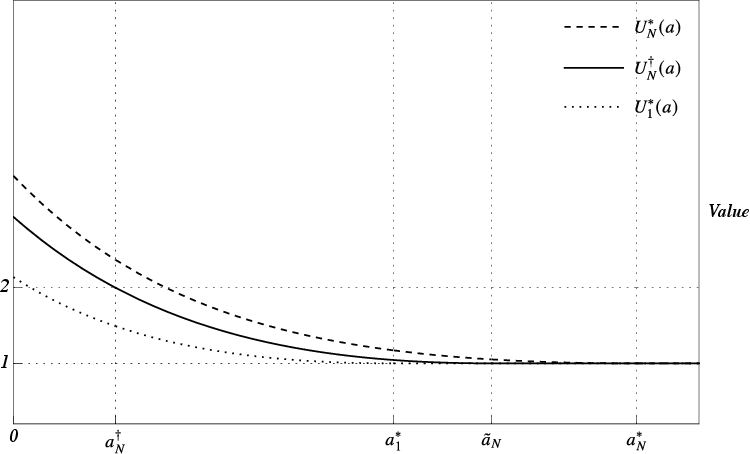}
    \caption{
    From top to bottom:
    average payoff \(U^*_N\) in the cooperative solution,
    common payoff \(U^\dagger_N\) in the symmetric equilibrium,
    and payoff \(U^*_1\) in the single-agent optimum.
    Parameter values: \(\rho = 2\), \(\theta = -1\), \(N=2\).}
    \label{fig:valueplotsym}
\end{figure}

As the outcomes are publicly observed
and newly developed technologies are freely available,
the players have incentives to free-ride.
Such a \emph{free-rider effect} becomes more pronounced through
the comparison between
the benefit-cost ratio of exploration
at the states of interior allocation
in the symmetric MPE
\[
    \beta(a, U^\dagger_N) = 1,
\]
and the one in the cooperative solution
\[
    \beta(a, U^*_N) = U^*_N(a)/N.
\]
Exploration in equilibrium thus requires
the benefit of exploration
to cover the cost,
whereas the efficient strategy entails exploration at states
where the cost exceeds the benefit,
as \(\beta(a,U^*_N) < 1\) whenever \(U^*_N(a) < N\).
Figure~\ref{fig:valueplotsym} illustrates the comparison
between the common payoff function in the symmetric equilibrium
and the cooperative solution in a two-player exploration game.

\subsection{Symmetric Equilibrium: Comparative Statics}

In this section, we examine the comparative statics 
of the symmetric equilibrium with respect to
the number of players \(N\).

\begin{corollary}
    \label{cor:cs-n}
    On \(\{\,N\geq 1\mid a^\dagger_N > 0\,\}\),
    which is the range of \(N\) for which
    the players' resource constraints are binding in the symmetric equilibrium,
    the stopping threshold \(\tilde{a}_N\)
    is strictly increasing in \(N\)
    and the common payoff function \(U^\dagger_N\)
    is weakly increasing in \(N\).
    Whereas on \(\{\,N\geq 1 \mid a^\dagger_N = 0\,\}\),
    both \(\tilde{a}_N\) and \(U^\dagger_N\)
    are constant over \(N\),
    and the equilibrium strategy \(k^\dagger_N\) is weakly decreasing in \(N\).

\end{corollary}

In the binding case,
note that
\(k^\dagger_N(a)\) is not monotone in \(N\)
because the full-intensity threshold \(a^\dagger_N\) could be decreasing in \(N\).
This situation occurs when the free-rider effect outweighs the encouragement effect.
On the one hand,
extra encouragement brought by additional players
raises the stopping threshold \(\tilde{a}_N\).
On the other hand,
the increased free-riding incentives due to extra players
tighten the requirement \(u > N\) for binding resource constraints,
which enlarges the region \((a^\dagger_N, \tilde{a}_N)\) of interior allocation.
The total effect of increasing \(N\)
on the intensity of exploration
is determined by these two competing forces
and hence is not monotone in \(N\).

In the non-binding case,
as \(N\) increases,
each player adjusts their individual intensity of exploration downward,
maintaining the same equilibrium payoff.
The incentive to free-ride in such a situation is so strong that it completely offsets further encouragement brought by additional players.
Even the overall intensity of exploration \(Nk^\dagger_N\) is decreasing in \(N\) 
for the gaps in \([0,\tilde{a}_N)\).
Thus, free-riding slows down exploration considerably.
In the worst scenario,
the overall intensity when the gap is zero
could even be lower than that in the single-agent problem.

Also, note that whenever
the resource constraints are not binding,
the full-intensity threshold \(a^\dagger_N\)
remains constant at zero for any further increase in \(N\)
because \(k^\dagger_N\) would be even lower.
Therefore,
if \(a^\dagger_N\) ever hits zero as \(N\) goes up,
the resource constraints in the symmetric equilibrium
remain non-binding for any larger \(N\).

As we have seen,
depending on whether or not the resource constraints are binding
in the symmetric equilibrium,
a larger team size
can lead to qualitatively difference outcomes.
If the resource constraints are not binding in equilibrium,
any extra resource brought by additional players
translate entirely to free-riding,
which results in a highly inefficient outcome in a large team.
The question then naturally arises
of whether the resource constraints would ever fail to be binding
in the symmetric equilibrium
as \(N\) goes up.
Or, conversely,
would the encouragement effect eventually overcome the free-rider effect?\footnote{
    The purpose of classifying the symmetric MPE into binding and non-binding cases
    is to help \emph{describe} the comparative statics.
    We do not intend to suggest 
    this classification of equilibria in a large team \emph{determines}
    the prevalence of the encouragement or free-rider effect.
    On the very contrary, it is the \emph{consequence} of the relative strength between these two forces.
}

To investigate this question,
we now examine the effect on the symmetric equilibrium
as \(N\) increases toward infinity,
while keeping the other parameters fixed.
For ease of exposition, 
we allow \(N\geq 1\) to take non-integral values
for the remainder of this section.

\begin{corollary}
    \label{cor:asymp-n}

    If \(W\) yields IRC and \(r < \hat{r} \coloneqq 
    (\sigma\theta)^2/(2(\theta-\ln(1+\theta)))\),\footnote{
        Note that \(\hat{r}\) is well defined only if \(W\) yields IRC.
    }
    then we have
    \(a^\dagger_N\to +\infty\)
    and
    \(U^\dagger_N\to +\infty\)
    as \(N\to 1/(\rho(1+\theta))\);
    otherwise,
    we have
    \(a^\dagger_N = 0\) for sufficiently large \(N\),
    and
    \(\lim U^\dagger_N(a) < \lim U^*_N(a)\) for each \(a\geq 0\)
    as \(N\to+\infty\).\footnote{
        Here,
        we allow Assumption~\ref{asm:main-assumption} to be violated as \(N\to+\infty\).
    }

\end{corollary}

The free-rider effect and the encouragement effect in our model
are two competing forces
shared in several models in the literature on strategic experimentation (e.g., \textcite{BoltonHarris:1999, KellerRady:2010,KellerRady:2015}).
In the symmetric equilibrium of the Brownian model of \textcite{BoltonHarris:1999},
the free-rider effect eventually dominates the encouragement effect
as team size grows.
This is not necessarily the case here.

If the technological landscape yields DRC,
then the marginal welfare improvement is certainly diminishing asymptotically
with respect to the team size.
Unsurprisingly, like the Brownian model,
the marginal encouragement effect yields to
the marginal free-rider effect,
as the latter does not abate as the team gets larger.

\begin{figure}[htb]
    \centering
    \begin{subfigure}[b]{0.48\textwidth}
        \centering
        \includegraphics[width=\textwidth]{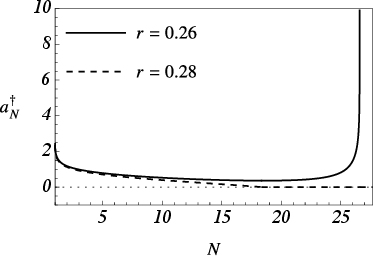}
        \caption{Full-intensity threshold \(a^\dagger_N\)}
        \label{fig:full-intensity-cutoff-n}
    \end{subfigure}
    \hfill
    \begin{subfigure}[b]{0.48\textwidth}
        \centering
        \includegraphics[width=\textwidth]{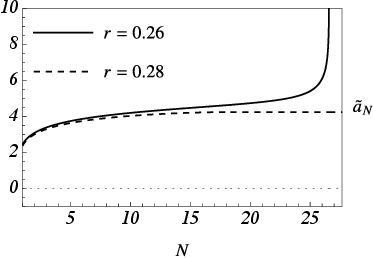}
        \caption{Stopping threshold \(\tilde{a}_N\)}
        \label{fig:stopping-cutoff-n}
    \end{subfigure}

       \caption{
           Full-intensity  thresholds
           and the stopping thresholds in the symmetric equilibrium
           (\(\theta = -0.09, \sigma = \sqrt{2}\))
           for different discount rate \(r\).
           If the players are sufficiently patient (solid curves),
           resource constraints are binding
           (\(a^\dagger_N > 0\)) for all \(N\).
           Otherwise (dashed curve),
           resource constraints are not binding
           (\(a^\dagger_N = 0\)) for sufficiently large \(N\).
        }
       \label{fig:cs-n}
\end{figure}

However,
when \(W\) does not lead to DRC,
the innovation effect introduced in Section~\ref{sec:cooperative-problem}
can make a difference.
Stemming from the encouragement effect,
the future exploration from an additional player
encourages everyone to become more ambitious,
which then leads to the emergence of more advanced technologies in the long run.
This novel innovation effect in our model in turn 
motivate the players to explore,
reinforcing the encouragement effect,
and vice versa.
Therefore, exploration over a non-DRC technological landscape allows the encouragement effect to prevail,
which is exhibited by the unlimited expansion of the full-intensity region, as illustrated in Figure~\ref{fig:cs-n}.

Even so,
collective exploration over a CRC or IRC landscape
is only necessary for the prevalence of the encouragement effect,
but not sufficient.
The returns to cooperation
determine how likely or how advanced the technologies are expected to be developed,
but the timing of their availability also matters.
Naturally,
players' patience plays a role:
The prevalence of the encouragement effect
requires both an IRC technological landscape
and sufficiently patient players,
as stated in Corollary~\ref{cor:asymp-n}.
The innovation effect from the exploration over a CRC landscape,
or an IRC landscape with impatient players,
fails to boost the encouragement effect up to the magnitude required for overcoming the free-rider effect.
In such cases,
as team size grows to infinity,
the full-intensity region vanishes
and the equilibrium payoff functions stay bounded,
standing in marked contrast to the cooperative solution
in which payoffs grow without bound.

Corollary~\ref{cor:asymp-n} 
highlights the key role of innovation
in strategic learning and experimentation,
which has been largely overlooked in the literature despite its importance.
The absence of technological advancements is partly responsible 
for the prevalence of free-riding in two-armed bandit models.
Our result suggests that innovation is an essential element toward
understanding the incentives
for experimentation and technology development
in dynamic and strategic environments.

\section{Asymmetric Equilibria and Welfare Properties}

Note that in the symmetric equilibrium,
the intensity of exploration dwindles down to zero
as the gap approaches the stopping threshold.
As a result,
the threshold is never reached
and exploration never fully stops.
This observation suggests that welfare can be improved if the players take turns
between the roles of explorer and free-rider,
keeping the intensity of exploration bounded away from zero 
until all exploration stops.
In this section, we investigate this possibility
by constructing a class of asymmetric Markov perfect equilibria.

\subsection{Construction of Asymmetric Equilibria}

Our construction of asymmetric MPE
is based on the idea of the asymmetric MPE proposed in \textcite{KellerRady:2010}.
We let the players adopt the common actions
in the same way as in the symmetric equilibrium
whenever the resulting average payoff is high enough
to induce an overall intensity of exploration greater than one,
and let the players take turns exploring at less promising states
in order to maintain the overall intensity at one.
Such alternation between the roles of explorer and free-rider
leads to an overall intensity of exploration
higher than in the symmetric equilibrium,
yielding higher equilibrium payoffs.

In what follows,
we briefly address the two main steps in our construction.
In the first step, we construct the average payoff function \(\bar{u}\).
We let \(\bar{u}\) solve the same ODE \(\beta(a, u) = \max\{u(a)/N, 1\}\)
as the common payoff function in the symmetric equilibrium
whenever \(u > 2 - 1/N\),
which ensures the corresponding overall intensity
is greater than one.
Whenever \(1 < u < 2 - 1/N\),
we let \(\bar{u}\) solve the ODE \(u(a) = 1 - 1/N +\beta(a, u)\),
which is the ODE for the average payoff function among \(N\) players
associated with an overall intensity \(K = 1\).
The boundary conditions for the average payoff function \(\bar{u}\),
namely the smooth pasting condition at the stopping threshold
and the normal reflection condition~\eqref{eq:normal-reflection},
are identical to the conditions in Lemma~\ref{lem:char-mpe},
simply because those conditions remain unchanged
after taking the average.
The unique solution of class \(C^1(\R_{++})\)
to the ODE above
serves as the average payoff function,
which also gives thresholds \(a^\flat > a^\sharp \geq 0\)
such that \(\bar{u} = 1\) on \([a^\flat, +\infty)\),
\(1 < \bar{u} < 2 - 1/N\) on \((a^\sharp, a^\flat)\)
and \(\bar{u} > 2 - 1/N\) on \([0, a^\sharp)\).
In the second step,
equilibrium-compatible actions
are assigned to each player.
On \([0, a^\sharp)\),
if it is nonempty,
we let the players adopt the common action
\(k_n(a) = \min\{(\bar{u}(a) - 1)/(N - 1), 1\}\)
in the same way as in the symmetric equilibrium.
On \([a^\sharp, a^\flat)\),
players alternate between
the roles of explorer and free-rider
so as to keep the overall intensity at one.
We first split \([a^\sharp, a^\flat)\) into subintervals
in an arbitrary way
and then meticulously choose the switch points of their actions,
so that all individual payoff functions
have the same values and derivatives
as the average payoff function
at the endpoints of these subintervals.\footnote{
    In fact, this technique can be used to construct asymmetric equilibria
    with strategies that take values in \(\{0,1\}\) only,
    which are referred to as simple equilibria in \textcite{KellerRadyCripps:2005}.
    However, it is not clear whether such equilibria
    achieve higher average payoffs than the symmetric equilibrium.
}
Lastly, our characterization of MPE in Lemma~\ref{lem:char-mpe}
confirms that the assigned action profile is compatible with equilibrium.
We leave the method for choosing the switch points
and further details to Appendix~\ref{sec:proofs-asym-mpe}.

\begin{figure}[htb]
    \centering
    \includegraphics[width=\textwidth]{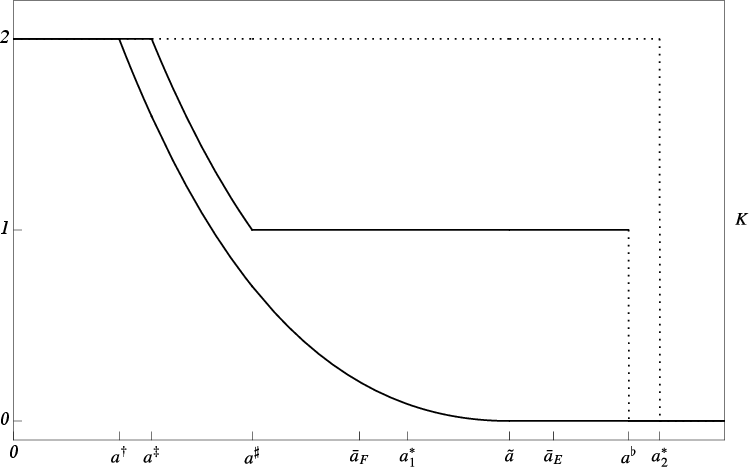}
    \caption{From top to bottom:
    intensity of exploration in the cooperative solution,
    the asymmetric equilibria,
    and the symmetric equilibrium
    (\(\rho = 2\), \(\theta = -1\), \(N=2\)).}
    \label{fig:kplotasym}
\end{figure}

\begin{figure}[htb]
    \centering
    \includegraphics[width=\textwidth]{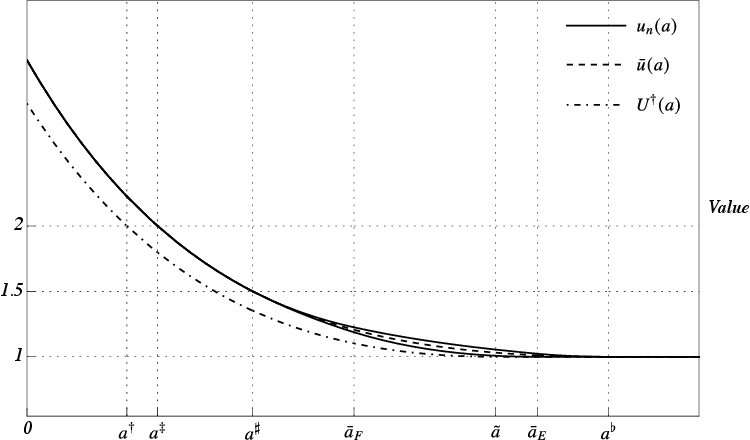}
    \caption{
        Average payoff and possible individual payoffs
        in the best two-player asymmetric equilibria,
        compared to the common payoff in the symmetric equilibrium
        (\(\rho = 2\), \(\theta = -1\), \(N=2\)).
    }
    \label{fig:valueplotasym}
\end{figure}

For \(N = 2\),
Figure~\ref{fig:kplotasym}
illustrates the intensity of exploration
in the asymmetric MPE, compared with the symmetric equilibrium.
The resource constraints are binding for small gaps in the depicted equilibria,
but this may not be the case for different parameters.
For example,
if the players are too impatient,
the average payoff function could be bounded by \(2-1/N\) from above,
resulting in an intensity of exploration equal to 1
over the entire region \([0, a^\flat)\).
The states \(\bar{a}_{F}\) and \(\bar{a}_{E}\) in the figure
demarcate the switch points
at which these two players swap roles
when they take turns exploring on \([a^\sharp, a^\flat)\).
The volunteer explores on
\([a^\sharp, \bar{a}_{F}) \cup [\bar{a}_{E}, a^\flat)\),
whereas the free-rider explores on \([\bar{a}_{F}, \bar{a}_{E})\).
These switch points are chosen in a way that
ensures their individual payoff functions
are of class \(C^1(\R_{++})\)
and coincide on \([0, a^\sharp)\).

Figure~\ref{fig:valueplotasym} illustrates
the associated average payoff function (dashed curve)
and the individual payoff functions (solid curves) that can arise in the equilibria
in a two-player game,
compared with the common payoff function
in the symmetric equilibrium (solid dotted curve).
Note that the payoff function of the volunteer
is strictly higher than that of the free-rider
at the states immediately to the left of
the stopping threshold \(a^\flat\).
In fact,
the free-rider has a payoff equal to 1 on \([\bar{a}_{E}, a^\flat)\).
This observation stands in marked contrast to the models
in \textcite{KellerRadyCripps:2005,KellerRady:2010},
where the volunteer is worse off in this region.
The intuition behind this feature
is similar to that in Proposition~\ref{prop:everyone-explores}.
A kink has to be created at \(a^\flat\)
in the free-rider's payoff function
in order to attain a higher payoff than the volunteer's
at states immediately to the left of \(a^\flat\),
because the free-rider's ODE \(u(a) = 1 + \beta(a, u)\) must be satisfied.\footnote{
    Even though the free-rider has a payoff equal to 1
    around \(a^\flat\),
    she still benefits from free-riding in this region.
    This benefit, however, is offset by the
    relatively high burden of exploration effort she must bear in equilibrium
    at more promising states to reward the volunteer.
}
In such a case, the free-rider has a strict incentive
to take over the role of volunteer
to kickstart the exploration process at a larger gap.
Therefore, in equilibrium,
the volunteer must be compensated
for acting as a lone explorer at less promising states
by bearing relatively less burden at more promising states.

For arbitrary \(N\), 
we have the following result.
\begin{proposition}[Asymmetric MPE]
    \label{prop:asymmetric-mpe}
    The \(N\)-player exploration game admits
    Markov perfect equilibria with thresholds
    \(0\leq a^\ddag_N \leq a^\sharp_N < a^\flat_N < a^*_N\),
    such that
    on \([0, a^\sharp_N]\),
    the players have a common payoff function;
    on \([0, a^\ddag_N]\),
    all players choose exploration exclusively;
    on \((a^\ddag_N, a^\sharp_N)\),
    the players allocate a common interior fraction
    of the unit resource to exploration,
    and this fraction decreases in the gap;
    on \([a^\sharp_N, a^\flat_N)\),
    the intensity of exploration equals 1
    with players taking turns exploring
    on consecutive subintervals;
    on \([a^\flat_N, +\infty)\),
    all players choose exploitation exclusively.
    The intensity of exploration is continuous in the gap on \([0, a^\flat_N)\).
    The average payoff function is strictly decreasing on \([0, a^\flat_N]\),
    once continuously differentiable on \(\R_{++}\),
    and twice continuously differentiable on \(\R_{++}\)
    except for the cutoff \(a^\flat_N\).
    On \([0, a^\flat_N)\),
    the average payoff is higher than in the symmetric equilibrium,
    and \(a^\flat_N\) lies to the right of the threshold \(\tilde{a}_N\)
    at which all exploration stops in that equilibrium.
\end{proposition}

\subsection{Welfare Results}

For \(N\geq 3\),
further improvements can be easily achieved
by letting the players take turns exploring,
maintaining the intensity of exploration
at \(K\) whenever \(K < u < K + 1 - K/N\)
for all \(K\in\{1,\ldots, N\}\),
rather than for \(K=1\) only as in the Proposition above.
However, it is not clear whether such improvements achieve
the highest welfare among all MPE
of the \(N\)-player exploration game.
For \(N=2\),
the asymmetric equilibria of Proposition~\ref{prop:asymmetric-mpe}
are the best among all MPE.

\begin{proposition}[Best MPE for \(N=2\)]
    \label{prop:bound-two-player-mpe}
    The average payoff in any Markov perfect equilibrium
    of the two-player exploration game
    cannot exceed the average payoff in the equilibria
    of Proposition~\ref{prop:asymmetric-mpe}.
\end{proposition}

In the construction of the asymmetric MPE depicted in Figure~\ref{fig:kplotasym},
the interval \([a^\sharp, a^\flat)\) is not split into subintervals.
This assertion can be confirmed by the observation from Figure~\ref{fig:valueplotasym}
that the players' payoff functions match values
only at the endpoints of \([a^\sharp, a^\flat]\),
not in the interior.
Our construction allows an arbitrary partition on \([a^\sharp, a^\flat)\)
during the splitting procedure,
thus a trivial partition of \([a^\sharp, a^\flat)\),
as in Figure~\ref{fig:kplotasym}, suffices.
A finer partition,
however,
produces equilibria in which the players exchange roles more often,
which allows them to share the burden of exploration more equally.
Sufficiently frequent alternation of roles on \([a^\sharp, a^\flat)\)
guarantees each player a payoff close enough to the average payoff
and thus yields a Pareto improvement over the symmetric equilibrium.\footnote{
    The payoffs of both players in the asymmetric MPE depicted in Figure~\ref{fig:valueplotasym}
    are higher than in the symmetric equilibrium
    on \([0, \bar{a}_E)\);
    however, this might not be the case in general
    when the trivial partition of \([a^\sharp, a^\flat)\) is used in the construction,
    as \(\bar{a}_{E}\) could lie on the left of \(\tilde{a}_N\).
}
\begin{proposition}[Pareto Improvement over the Symmetric MPE]
    \label{prop:pareto}
    For any \(\epsilon > 0\),
    the \(N\)-player exploration game admits
    Markov perfect equilibria as in Proposition~\ref{prop:asymmetric-mpe}
    in which each player's payoff exceeds the symmetric equilibrium payoff 
    on \([0, a^\flat_N - \epsilon]\).
\end{proposition}

Recall that
the common payoff function in the symmetric equilibrium
is constant over the team size
when the resource constraints are not binding.
By contrast,
it can be shown that
the average payoff always increases in the number of players 
in the asymmetric MPE in which the players take turns exploring
at states immediately to the left of the stopping cutoff.
This is because the burden of keeping the overall intensity at one
at these states
can be shared among more players in larger teams.
As a result,
the players would be able to exploit more often on average,
which in turn encourages them to explore at less promising states.
Nevertheless,
the free-rider effect can still prevail
in a large team
when the players are not sufficiently patient,
just like in the symmetric equilibrium.
It can be shown that
the results regarding the average payoff
and the full-intensity threshold in Corollary~\ref{cor:asymp-n}
extend to the asymmetric MPE of Proposition~\ref{prop:asymmetric-mpe}
with a larger \(\hat{r}\).

\section{Discussion}

In this section,
we discuss our modeling assumptions
and assess the extent to which our results rely on them.
We argue that
our assumptions establish a parsimonious environment,
suggesting that our findings regarding the prevalence of the encouragement effect remain robust across various plausible extensions.

\paragraph*{Payoffs.}
We have assumed that
players receive payoffs only from exploitation,
which serves to highlight
the innovation-driven motives for exploration.
This assumption deviates from
the literature on strategic experimentation such as \textcite{BoltonHarris:1999,KellerRadyCripps:2005}, and the literature on spatial experimentation such as \textcite{Callander:2011,GarfagniniStrulovici:2016},
in which the players also receive payoffs directly from experimentation.
Notably, allowing the players to benefit from exploration per se
barely changes our results.
For example, 
suppose in addition to the flow payoffs from exploitation,
the players also receive a flow payoff of \(k_{n,t}\exp(W(X_t))\dd{t}\) from exploration.
In such a case, the ``gap'' still serves as a state variable,
but some of the closed-form representations
in our results may no longer be attainable.
The players would then strictly prefer exploration over exploitation
when the gap is sufficiently close to zero,
as exploration offers a positive option value
in addition to a flow payoff
that is nearly identical to that of exploitation.
In other words,
not only does the prospect of technological advancements motivate the players to explore,
but also exploration per se.
As a result,
the players would 
have a stronger incentive to explore,
and thus the encouragement effect would
prevail under weaker conditions.

We have also assumed that
the qualities of the technologies contribute
exponentially to the payoffs from exploitation.
This assumption aligns with the exponential growth of \emph{total factor productivity},
commonly assumed in macroeconomic growth literature dating back to \textcite{Solow:1956}.
Some empirical observations, such as Moore's law (doubling of transistors on integrated circuits every two years),
are in line with exponential growth,
while others do not (see, e.g., \textcite{Philippon:2022}).
We make the exponential growth assumption mainly for tractability,
because it helps reduce the dimension of the state variable.
This simplification can also be achieved by choosing
a factor depending linearly on the qualities of the technologies
(i.e., adopting the best-known technology delivers
a flow payoff of \((1-k_{n,t})S_t\dd{t}\)
instead of \((1-k_{n,t})\exp(S_t)\dd{t}\)),
but with the underlying landscape represented by a \emph{geometric} Brownian motion.
All our results continue to hold in such an equivalent formulation,
with the ratio \(S_t / W(X_t)\), or its monotone transformation,
such as \(\ln(S_t) - \ln(W(X_t))\),
serving as a one-dimensional state variable.

We suspect that it is inevitable to resort to a two-dimensional state variable such as \((w, s) = (W(X_t), S_t)\) for other function forms of the flow payoff.
The challenge mainly arises from the lack of homogeneity in payoff functions stated in Lemma~\ref{lem:homogeneity}.
Without homogeneity,
it would be difficult to pin down the equilibrium payoff functions.
The analysis for each given \(s\) remains similar to the current setting,
but the equilibrium strategies could be hard to analyze and interpret
if the strategies are unrestricted along the \(s\)-coordinate.
We also suspect that
if the order of growth is lower than the exponential rate,
the encouragement effect would be unable to overcome the free-rider effect.
Moreover, reward and punishment become possible
by conditioning actions on the highest-known quality \(s\),
which probably leads to more efficient outcomes as in \textcite{HornerKleinRady:2021}.
In that paper, they demonstrate that inefficiencies disappear entirely in a class of non-Markovian equilibria in a rich environment that encompasses the Brownian and Poisson models.
Since our setting lies outside of their environment,
it remains an open question whether the insight from their constructive proofs can be applied to our setting to achieve full efficiency,
complementary to the positive results that we obtained here by focusing only on MPE.

\paragraph*{Exploration.}
The scope of experimentation is certainly limited in our model:
Players do not have complete freedom to choose where to explore.
\textcite{GarfagniniStrulovici:2016}
allow the players to experiment with any technologies,
but radical experimentation,
which involves exploring technologies far away from the feasible ones,
is assumed to be more expensive.
Exploration in our model can be viewed as an extreme abstraction of their model,
where incremental experimentation is costless,
while radical experimentation comes at an infinite cost.
In practice,
such a limited experimentation scope may be more appropriate
in the context of technology development.
For example, pharmaceutical companies can easily test the efficacy of a medicine
once its formula is provided,
but creating the formula from scratch is nearly impossible.

It might also be reasonable to assume
that multiple research directions emerge during exploration,
allowing players to pursue them concurrently or switch direction if one proves fruitless.
Such a possibility is beyond the scope of this paper,
but it is expected to give rise to a stronger encouragement effect,
as restarting opportunities would raise the likelihood of innovation
and therefore the value of exploration as well.
\section{Concluding Remarks}

This paper introduces a novel and tractable framework
for examining the incentives 
of forward-looking agents
in knowledge creation.
We identify two key effects that shape the incentives for experimentation:
an encouragement effect,
unique to strategic and dynamic contexts,
and an innovation effect,
which is absent from the existing strategic bandit literature where innovation possibilities are often overlooked.
We demonstrate that the innovation effect,
stemming from the prospect of technological advancements,
can amplify the encouragement effect, thereby offsetting the free-rider problem prevalent in large teams.
Our analysis further illustrates
how these effects impact the trajectory of technological progress
and long-run outcomes.
The proposed model holds promise for future research,
with potential applications in dynamic games of innovation,
such as patent races.

\appendix
\appendixpage

\section{Explicit  Representation of the Symmetric MPE}
\label{sec:sym-mpe-expression}
\begin{corollary}
    \label{cor:symmetric-mpe}

    The explicit representation for
    the normalized payoff function \(U^\dagger_N\)
    in the unique symmetric equilibrium
    on \([a^\dagger_N, \tilde{a}_N)\) is given by
    \[
        U^\dagger_N(a) = 
        \begin{cases}
            1 + \frac{1}{2\rho}(\tilde{a}_N - a)^2, & \text{ if }\theta= 0,\\
            1 + \frac{1}{\rho\theta}\left(\tilde{a}_N - a+\frac{1}{\theta}\left(\me^{-\theta(\tilde{a}_N - a)}-1\right)\right), & \text{ if }\theta\neq 0.
        \end{cases}
    \]

    If \(a^\dagger_N > 0\), \(U^\dagger_N\) on \([0, a^\dagger_N)\) is given by
    \[
        U^\dagger_N(a) = \frac{N}{\gamma_2 - \gamma_1}
        \left(
            (\gamma_2 + \iota) \me^{-\gamma_1(a^\dagger_N - a)}
            - (\gamma_1 + \iota) \me^{-\gamma_2(a^\dagger_N - a)}
        \right),
    \]
    with \(\gamma_1 < \gamma_2\) being the roots of the equation \(\gamma(\gamma - \theta) = 1/(N\rho)\)
    and
    \(\iota > 0\) being
    \[
        \iota \coloneqq \begin{cases}
            \frac{1}{N\rho \theta}\left(1+ W_{0}\left(-\exp(-1 - (N-1)\rho\theta^2)\right)\right), & \text{ if }\theta > 0, \\
            \sqrt{\frac{2}{N\rho}(1-1/N)}, &  \text{ if }\theta = 0, \\
            \frac{1}{N\rho\theta}\left(1 + W_{-1}\left(-\exp(-1 - (N-1)\rho\theta^2)\right)\right), & \text{ if }\theta < 0,
    \end{cases}
    \]
    where \(W_0\) and \(W_{-1}\) denote the real branches of the Lambert \(W\) function.\footnote{
        The Lambert \(W\) function maps each \(x \geq -1/\me\) to the solutions of the equation \(y \me^y = x\).
        It has two real branches.
        The value of the principal branch \(W_0(x)\) denotes the unique solution \(y\)
        for which \(y\geq -1\),
        whereas the value of the branch \(W_{-1}(x)\) denotes the unique solution \(y\)
        for which \(y < -1\).
    }

    If \(\iota < 1\),
    then the equilibrium belongs to the binding case.
    The full-intensity threshold is given by
    \[
        a^\dagger_N
        = \frac{1}{\gamma_2 - \gamma_1}\left(
            \ln(\frac{1 + \gamma_2}{\iota + \gamma_2})
            -\ln(\frac{1 + \gamma_1}{\iota + \gamma_1})
        \right),
    \]
    and the stopping threshold is given by
    \[
        \tilde{a}_N = a^\dagger_N + N\rho(\iota + \theta(1-1/N)).
    \]

    If \(\iota \geq 1\),
    then the equilibrium belongs to the non-binding case with the full-intensity threshold \(a^\dagger_N = 0\),
    whereas the stopping threshold \(\tilde{a}_N\) is given by
    \begin{align*}
        \tilde{a}_N &= \begin{cases}
            \frac{1}{\theta}\left(1+\theta-\theta^2\rho + W_{0}\left(-(1+\theta)\me^{-1-\theta+\theta^2\rho}\right)\right), & \text{ if }\theta < 0, \\
            1-\sqrt{1-2\rho}, & \text{ if }\theta = 0, \\
            \frac{1}{\theta}\left(1+\theta-\theta^2\rho + W_{-1}\left(-(1+\theta)\me^{-1-\theta+\theta^2\rho}\right)\right), & \text{ if }\theta > 0. \\
    \end{cases}
    \end{align*}
\end{corollary}

\begin{proof}
    These expressions follow from Lemma~\ref{lem:char-mpe} and explicit calculations.
\end{proof}

\section{Properties of Payoff Functions}
\label{sec:proofs-ppt-payoff-functions}
\begin{lemma}[Homogeneity]
    \label{lem:homogeneity}
    Player \(n\)'s payoff function
    for an \(s\)-invariant Markov strategy profile
    \(\bm{k}\in \mathcal{K}^N\)
    can be written as
    \(v_n(\, a,s \mid \bm{k} \,) = \me^s v_n(\, a, 0 \mid \bm{k}) \,\).
\end{lemma}

\begin{proof}
    At state \((a,s)\),
    player \(n\)'s payoff associated to \(\bm{k}\)
    is given by
    \begin{align*}
        v_n(\, a,s \mid \bm{k} \,) &= {\E}\left[
                \int_0^\infty r \me^{-rt} (1-k_n(A_t)) \me^{S_t} \dd{t}
                \;\middle\vert\; A_0=a, S_0 = s
        \right]\\
        &=  \me^s {\E}\left[
                \int_0^\infty r \me^{-rt} (1-k_n(A_t)) \me^{S_t - s} \dd{t}
            \;\middle\vert\; A_0=a, S_0 = s
        \right] \\
        &=  \me^s {\E}\left[
                \int_0^\infty r \me^{-rt} (1-k_n(A_t)) \me^{S_t} \dd{t}
                \;\middle\vert\; A_0=a, S_0 = 0
        \right] \\
        &= \me^s v_n(\, a, 0 \mid \bm{k} \,),
    \end{align*}
    where the second-to-last equality comes from
    the Markov property of the diffusion process
    \(\{A_t\}_{t\geq 0}\).
\end{proof}

\begin{lemma}[Smoothness]
    \label{lem:smooth-payoff}
    Given an \(s\)-invariant Markov strategy profile
    \(\bm{k}\in \mathcal{K}^N\),
    for each \(s\),
    player \(n\)'s payoff function \(u_n(\, \cdot \mid \bm{k} \,)\)
    is 
    \begin{enumerate}[label=(\roman*)]
        \item \(C^1\) and piecewise \(C^2\)
        on \((a_L, a_R)\subset \R_+\)
        if the intensity of exploration \(K(a)\) is bounded away from zero on \((a_L, a_R)\);
        \label{item:lem-smooth-payoff-1}
        \item \(C^2\) on \((a_L, a_R)\)
        if both \(K(a)\) and \(k_n(a)\) are continuous on \((a_L, a_R)\) in addition to the condition above.
        \label{item:lem-smooth-payoff-2}
    \end{enumerate}
\end{lemma}

\begin{proof}
    Here, we present the proof of part~\ref{item:lem-smooth-payoff-1},
    where it is not necessary for \(\bm{k}\) to be continuous.
    Part~\ref{item:lem-smooth-payoff-2} is a standard result and thus the proof is omitted.\footnote{See, e.g., the proof of Theorem 4.5 in \textcite{Durrett:1996}, p.\ 226.}
    
    We first show that
    given the payoffs \(u_L\coloneqq u_n(\, a_L \mid \bm{k} \,)\) 
    and \(u_R\coloneqq u_n(\, a_R\mid \bm{k} \,)\)
    at the boundaries,
    if there exists a function \(v\) of class \(C^1\) and piecewise \(C^2\)
    on \((a_L, a_R)\)
    which solves the two-point boundary value problem (BVP)
    \begin{align*}
        \Lgen v(a) - r v(a) &= - f(a),\\
        v(a_L) &= u_L, \\
        v(a_R) &= u_R,
    \end{align*}
    with \(\Lgen v(a) \coloneqq -\mu K(a) v'(a) + \frac{1}{2}\sigma^2 K(a) v''(a)\)
    and
    \(f(a)\coloneqq r (1-k_n(a))\me^s\),
    then we have \(v = u_n\) on \([a_L, a_R]\).
    We then argue that such a solution exists and the proof is complete.

    Suppose \(v\) is of class \(C^1\) and piecewise \(C^2\)
    and solves the above BVP.
    Denote the first time that \(A_t\) leaves \((a_L, a_R)\)
    by \(\tau \coloneqq \inf \{\, t > 0 \mid A_t \not\in (a_L, a_R) \,\}\),
    and note that it is a bounded stopping time.
    We can apply It\^{o}'s formula\footnote{
        It\^{o}'s formula applies to any function that is \(C^1\) and piecewise \(C^2\).
    }
    to \(\me^{-rT}v(A_T)\) and have
    \begin{align*}
        \me^{-r{(T\land\tau)}}v(A_T) - v(a)
        &= \int_0^{T\land\tau} \me^{-rt}(\Lgen v - rv)(A_t)\dd{t} 
        - \int_0^{T\land\tau} \me^{-rt} \sigma\sqrt{K(A_t)} v'(A_t)\dd{B_t}.
    \end{align*}
    We replace \(\Lgen v - r v\) with \(-f\)
    and rearrange to obtain
    \begin{align*}
        \me^{-r{(T\land\tau)}}v(A_T) - v(a) + \int_0^{T\land\tau} \me^{-rt}f(A_t)\dd{t} = -\int_0^{T\land\tau} \me^{-rt} \sigma\sqrt{K(A_t)} v'(A_t)\dd{B_t},
    \end{align*}
    which implies the left-hand side is a continuous martingale starting from 0.
    We can then apply the optional stopping theorem and have
    \begin{align*}
        v(a) &= \E_a\left[
            \int_0^\tau \me^{-rt}f(A_t)\dd{t} + e^{-r\tau}v(A_\tau)
        \right]\\
        &= \E_a\left[
            \int_0^\tau \me^{-rt}f(A_t)\dd{t}
        \right]
        + \E_a[e^{-r\tau}\ind_{\{A_\tau = a_L\}}]\, u_L
        + \E_a[e^{-r\tau}\ind_{\{A_\tau = a_R\}}]\, u_R \\
        &= u_n(\,a \mid \bm{k}\,),
    \end{align*}
    which shows that the solution to the BVP coincides with player \(n\)'s payoff function
    on \((a_L, a_R)\).

    We next prove the existence of a solution of class \(C^1\) and piecewise \(C^2\)
    to the above BVP,
    using the standard ``shooting method''
    (see, e.g., \textcite{StruloviciSzydlowski:2015}, p.\ 1044).
    Without loss of generality,
    we assume both \(K(a)\) and \(k_n(a)\)
    are Lipschitz-continuous on
    \((a_L, a_M)\cup(a_M, a_R)\).\footnote{
        Recall that we require \(k_n\), and hence also \(K\),
        to be right-continuous and piecewise Lipschitz-continuous.
        Here we assume both functions have only one possible discontinuity point at \(a_M\in(a_L, a_R)\).
    }
    Consider the initial value problem (IVP)
    \begin{align*}
        \Lgen v(a) - r v(a) &= -f(a), \\
        v(a_L) &= u_L, \\
        v'(a_L) &= 0,
    \end{align*}
    and the IVP with the homogeneous ODE
    \begin{align*}
        \Lgen v(a) - r v(a) &= 0, \\
        v(a_L) &= 0, \\
        v'(a_L) &= 1.
    \end{align*}
    Since \(K(a)\) is bounded away from zero,
    standard result (see, e.g., \textcite{Barbu:2016}, Theorem 2.4)
    shows that
    both problems admit unique solution defined on \((a_L, a_M)\)
    as long as the length of the interval \(\abs{a_L - a_M}\) is sufficiently small.
    By the same argument,
    we can then uniquely extend these solutions
    to the whole interval \((a_L, a_R)\),
    requiring them to be continuously differentiable at \(a_M\).
    Denote such an extension of the solution to the first IVP by \(v_f\),
    and to the homogeneous IVP by \(v_0\).
    Then \(v_f + \frac{u_R - v_f(a_R)}{v_0(a_R)}v_0\) is the solution
    of class \(C^1\) and piecewise \(C^2\) to the BVP
    as long as \(v_0(a_R)\neq 0\).

    Lastly, we argue that \(v_0(a_R)\neq 0\).
    Suppose by contradiction that \(v_0(a_R)=0\).
    Since \(v_0'(a_L) = 1 > 0\),
    \(v_0\) achieves its maximum at some \(\hat{a}\in(a_L, a_R)\)
    with \(v_0(\hat{a}) > 0\) and \(v_0'(\hat{a}) = 0\).
    As \(v_0\) solves \(\Lgen v - rv =  0\),
    we have
    \(L v(a) - r v(a) \to 0\)
    as \(a\to\hat{a}-\),
    which implies \(\lim_{a\to\hat{a}-} v_0''(a) = r v_0(\hat{a})/(\sigma^2 K(\hat{a})/2) > 0\),
    contradicting the fact that \(v_0\) attains its maximum at \(\hat{a}\).
\end{proof}

\begin{lemma}[Normal Reflection Condition and Feynman-Kac Equation]
    \label{lem:payoff-function-ppt-x}
    Given an \(s\)-invariant Markov strategy profile,
    the associated payoff function of each player
    satisfies Feynman-Kac equation~\eqref{eq:feynman-kac-x}
    at each state at which it is twice continuously differentiable.
    Moreover,
    if the intensity of exploration
    is bounded away from 0
    in some neighborhood of the state \(a = 0\),
    the payoff function also satisfies
    the normal reflection condition~\eqref{eq:normal-reflection}.
\end{lemma}

\begin{proof}
    A more general version of this lemma is provided in
    Lemma~\ref{lem:payoff-function-ppt-xs} below
    for Markov strategy profile \(\bm{k}\)
    with the state variable \((a,s)\).
    The normal reflection condition \((\pdv*{v_n}{a} + \pdv*{v_n}{s})(0+,s) = 0\)
    takes form of \(u_n(0) + u_n'(0+) = 0\)
    because of the homogeneity of the payoff functions
    for \(s\)-invariant strategies.
\end{proof}

Given Markov strategy profile \(\bm{k}\)
with the state variable \((a,s)\)
in which the strategies are not necessarily
best responses against each other,
let
\[
    {\Lgen}= \eta(a,s){\pdv{a}} + \frac{1}{2}\alpha^2(a,s){\pdv[2]{a}}
\]
with
\(\eta(a,s) = -\mu K(a,s)\)
and 
\(\alpha(a,s) = \sigma \sqrt{K(a,s)}\).
Denote by \(f(a,s) = r(1-k_n(a,s)) \me^s\)
the flow payoff that player \(n\) receives at state \((a,s)\),
and by \(v(a,s) = v_n(\, a,s \mid \bm{k} \,)\)
her payoff at state \((a,s)\).
\begin{lemma}[Normal Reflection Condition and Feynman-Kac Equation]
    \label{lem:payoff-function-ppt-xs}

    If \(a\mapsto v(a,s)\) is twice continuously differentiable at \(a\),
    then it satisfies the Feynman-Kac formula
    \begin{equation}
        \label{eq:feynman-kac-xs}
        r v(a,s) = {\Lgen} v(a,s) + f(a,s).
    \end{equation}
    Moreover,
    if \(v\)
    is twice continuously differentiable on \(\{a=0\}\),\footnote{
        Here we mean \(v:\R_{+}\times\R\to\R\)
        can be extended to a function
        that is twice continuously differentiable
        on some open set in \(\R^2\)
        containing \(\{a = 0\}\).
    }
    and \(K(a,s)\) is bounded away from 0
    on some open set containing \(\{a = 0\}\),
    then
    \(v\) satisfies the \emph{normal reflection} condition
    \begin{equation}
        \label{eq:normal-reflection-xs}
        \pdv{v(0+,s)}{a} + \pdv{v(0+,s)}{s} = 0.
    \end{equation}

\end{lemma}

\begin{proof}
    Given strategy profile \(\bm{k}\),
    consider player \(n\)'s continuation value at time \(T\),
    that is,
    \[
        v(A_T, S_T)
        = {\E}\left[
            \int_T^\infty \me^{-r(t-T)}f(A_t, S_t)\dd{t}
            \;\middle\vert\; \mathscr{F}_T
        \right],
    \]
    and her total discounted payoff at time 0,
    evaluated conditionally on the information available at time \(T\),
    that is,
    \begin{align*}
        Z_T &\coloneq {\E}\left[
            \int_0^\infty \me^{-rt}f(A_t, S_t)\dd{t}
            \;\middle\vert\; \mathscr{F}_T
        \right] \\
        &= {\E}\left[
            \int_T^\infty \me^{-rt}f(A_t, S_t)\dd{t}
            \;\middle\vert\; \mathscr{F}_T
        \right]
        + \int_0^T \me^{-rt}f(A_t, S_t)\dd{t} \\
        &= 
        \me^{-rT}v(A_T, S_T)  + \int_0^T \me^{-rt}f(A_t, S_t)\dd{t}.
    \end{align*}
    Note that \(\{Z_t\}_{t\geq 0}\) is a martingale.\footnote{
        Assumption~\ref{asm:main-assumption} guarantees
        all expectations in this proof are finite.
    }
    Indeed,
    for \(0\leq {T'}\leq T\),
    we have
    \begin{align*}
        {\E}[Z_T\vert \mathscr{F}_{T'}]
        &= {\E}\left[
            {\E}\left[
                    \int_0^\infty \me^{-rt}f(A_t, S_t)\dd{t}
                \;\middle\vert\;
                \mathscr{F}_T
            \right]
            \;\middle\vert\; \mathscr{F}_{T'}
        \right] \\
        &= {\E}\left[
            \int_0^\infty \me^{-rt}f(A_t, S_t)\dd{t}
            \;\middle\vert\; \mathscr{F}_{T'}
        \right] \\
        &= Z_{T'}.
    \end{align*}

    If It\^{o}'s formula can be applied to \(\me^{-r (T\land\tau)}v(A_{T\land\tau}, S_{T\land\tau})\) for some bounded stopping time \(\tau\),
    we would have
    \begin{align*}
        \me^{-r (T\land\tau)} v(A_{T\land\tau}, S_{T\land\tau}) - v(a,s)
        &= \int_0^{T\land\tau} \me^{-r t} ({\Lgen} v - r v)(A_t, S_t)\dd{t} \\
        &- \int_0^{T\land\tau} \me^{-rt} \alpha(A_t, S_t){\pdv{v}{a}}(A_t,S_t)\dd{B_t} \\
        &+ \int_0^{T\land\tau} \me^{-rt}
        \left(\pdv{v}{a} + \pdv{v}{s}\right)(A_t,S_t)\dd{S_t}.
    \end{align*}

    Because \(Z_T\) is a martingale starting from \(v(a,s)\),
    we can then conclude the process
    \begin{align*}
        & Z_{T\land\tau} - v(a,s) + \int_0^{T\land\tau} \me^{-rt}\alpha(A_t,S_t){\pdv{v}{a}}(A_t,S_t) \dd{B_t} \\
        &= \int_0^{T\land\tau} \me^{-rt}
        \left(
            ({\Lgen} v - r v + f)(A_t, S_t)
        \right)
        \dd{t}
        + \int_0^{T\land\tau} \me^{-rt}\left(\pdv{v}{a} + \pdv{v}{s}\right)(A_t,S_t)\dd{S_t}
    \end{align*}
    is a continuous martingale starting from 0.
    Observe that the right-hand side is a finite variation process.
    As a consequence, it must be 0 for all \(T\geq 0\) almost surely.

    Let \(\tau = \tau_1 \coloneqq \inf\{\, t > 0 \mid (A_t,S_t)\not\in O\times\{s\}\}\)
    for some neighborhood \(O\) of \(a\)
    in which \(a\mapsto v(a, s)\) is twice continuously differentiable.
    Feynman-Kac formula~\eqref{eq:feynman-kac-xs}
    then follows from the fact that
    the second integral on the right-hand side is 0 for all \(T\geq 0\)
    and thus the first integral is almost surely 0 for all \(T\geq 0\) as well.

    Similarly, consider
    \(\tau = \tau_2 \coloneqq \inf\{\, t > 0 \mid (A_t,S_t)\not\in [0, \epsilon_1)\times [s, s + \epsilon_2)\}\),
    for some \(\epsilon_1, \epsilon_2 > 0\)
    such that \(a \in [0, \epsilon_1)\),
    \(K(\cdot)\) is bounded away from zero on \([0, \epsilon_1)\),
    and \(v\) has a \(C^2\) extension on \((-\epsilon_1, \epsilon_1)\times(s-\epsilon_2, s+\epsilon_2)\) by our assumption.
    Moreover,
    note that the Lebesgue measure \(\dd{t}\)
    and the measure \(\dd{S_t}\) on \(([0,T\land\tau_2],\mathcal{B}([0,T\land\tau_2]))\)
    are mutually singular almost surely.
    Indeed, the set \(D\coloneq \{\,t\in[0,T\land\tau_2]:A_t = 0\,\}\)
    is a null set under Lebesgue measure \(\dd{t}\),
    and \([0,T\land\tau_2]\setminus D\) is a null set under measure \(\dd{S_t}\) almost surely.
    Because for all \(T\geq 0\), almost surely
    the sum of the integrals on the right-hand side is zero
    and \(\dd{t}\perp\dd{S_t}\),
    we know that for all \(T\geq 0\) each integrand
    is a.s.\ a.e.\ 0 on \([0,T\land\tau_2]\).\footnote{
        By a.e.\ 0 on \([0,T\land\tau_2]\)
        we mean the integrands are 0 on all non-null sets
        w.r.t.\ the corresponding measure.
    }
    This yields the normal reflection condition \eqref{eq:normal-reflection-xs}.\footnote{
        We require \(K(a,s) > 0\) in some neighborhood
        containing \(\{a = 0\}\),
        as otherwise all measurable subsets of \([0,T\land\tau_2]\)
        would be null sets w.r.t.\ \(\dd{S}_{t}(\omega)\),
        in which case we cannot conclude
        that \(v\) satisfies the normal reflection condition~\eqref{eq:normal-reflection-xs}.
    }
\end{proof}

\section{Equilibrium Characterization}
\label{sec:proofs-eq-char}

\subsection{Proof of Lemma~\ref{lem:char-mpe}}
Lemma~\ref{lem:char-mpe} follows from Lemma~\ref{lem:mpe-sufficiency} and~\ref{lem:mpe-necessity} below.

\begin{lemma}[Sufficiency]
\label{lem:mpe-sufficiency}
Given \(\bm{k}_{\neg n}\in\mathcal{K}^{N-1}\),
if \(u_n:\R_{+}\to\R\)
satisfies Condition~\ref{item:eq-char-1}--\ref{item:eq-char-4} in Lemma~\ref{lem:char-mpe},
then a piecewise right-continuous function \(k_n^*:\R_{+}\to[0,1]\)
which maximizes
the right-hand side of HJB equation~\eqref{eq:hjb}
at each continuity point of \(u_n''\)
is a best response against \(\bm{k}_{\neg n}\),
with
\(u_n\) being
the associated payoff function of player \(n\).
\end{lemma}
\begin{proof}

    The proof is a standard verification argument.
    See, e.g.,\ \textcite{FlemingSoner:2006}, Theorem III.9.1.
    The role played by the linear growth condition in a standard proof
    is instead played by Assumption~\ref{asm:main-assumption}.

    Given \(\bm{k}_{\neg n}\in\mathcal{K}^{N-1}\),
    for any admissible control process
    \(k_n = \{k_{n,t}\}_{t\geq 0}\in \mathcal{A}\),
    let
    \[
        {\Lgen} \coloneqq \eta(a,s,k_{n,t}){\pdv{a}} + \frac{1}{2}\alpha^2(a,s,k_{n,t}){\pdv[2]{a}},
    \]
    with
    \(\eta(a,s,k) = -\mu (k + K_{\neg n}(a))\)
    and 
    \(\alpha(a,s,k) = \sigma\sqrt{k + K_{\neg n}(a)}\).
    Let \(f(a,s,k) = r (1 - k) \me^s\),
    and consider \(v(a,s) = \me^s u_n(a)\),
    where \(u_n:\R_{+} \to \R\)
    satisfies Conditions~\ref{item:eq-char-1}--\ref{item:eq-char-4} in Lemma~\ref{lem:char-mpe}.

    Usually,
    applying It\^{o}'s formula to \(v\) at \(a > 0\)
    requires \(u_n\in C^2(\R_{++})\).
    However, It\^{o}'s formula is still valid for \(C^1\) functions
    with absolutely continuous derivatives,
    which is satisfied by \(u_n\) under Conditions~\ref{item:eq-char-1} and~\ref{item:eq-char-2}.\footnote{
        See, e.g.,\ \textcite{Chung:2014}, Remark 1, p.\ 187;
        \textcite{RogersWilliams:2000}, Lemma IV.45.9, p.\ 105;
        or \textcite{StruloviciSzydlowski:2015}, footnotes 73 and 78.
        \label{fn:ito-c1}
    }

    By applying It\^{o}'s formula to \(\me^{-r T} v(A_T, S_T)\),\footnote{
        Note that \(S_t\) is a finite variation process
        and hence both the quadratic variation \(\langle S, S \rangle_t\)
        and the covariation \(\langle A, S \rangle_t\) are identically zero.
        See \textcite{Shreve:2004}, Section 7.4.2
        for a non-technical treatment.
        \label{fn:ito}
    }
    for fixed \(0 < T < \infty\) we have
    \begin{align*}
        \me^{-r T} v(A_T, S_T) - v(a,s)
        &= \int_0^T \me^{-r t} ({\Lgen} v - r v)(A_t, S_t)\dd{t} \\
        &- \int_0^T \me^{-r t} \alpha(A_t, S_t, k_{n,t}){\pdv{v}{a}}(A_t,S_t)\dd{B_t}
        \\
        &+\int_0^T \me^{-r t} \left(\pdv{v}{a} + \pdv{v}{s}\right)(A_t, S_t)\dd{S_t}.
    \end{align*}

    Since \(\dd{S_t} = 0\) whenever \(A_t > 0\),
    and by Condition~\ref{item:eq-char-3} we have
    \(\left(\pdv*{v}{a} + \pdv*{v}{s}\right)(A_t, S_t) = 0\)
    when \(A_t = 0\),
    the last term is identically zero.
    Moreover, because \(\alpha(A_t, S_t, k_{n,t})\)
    and \(\pdv*{v}{a}\) are bounded,
    the second term has mean zero.
    Taking expectations on both sides,
    for fixed \(0 < T < \infty\)
    we have Dynkin's formula
    \[
    v(a,s)
    = -{\E}_{as}\left[
        \int_0^T \me^{-r t}({\Lgen} v - r v)(A_t, S_t) \dd{t}
    \right]
        + \me^{-r T}{\E}_{as}[v(A_T,S_T)],
    \]
    where the expectations on the right-hand side are finite for each \(T < \infty\).

    Note that HJB equation~\eqref{eq:hjb} implies
    \[
        r v(a,s) \geq f(a,s,k_{n,t}) + {\Lgen} v(a,s),
    \]
    and therefore we have
    \begin{equation}
        \label{eq:hjb-verification}
        v(a,s)
        \geq {\E}_{as}\left[
            \int_0^T \me^{-r t}f(A_t, S_t, k_{n,t}) \dd{t}
        \right]
            + \me^{-r T}{\E}_{as}[v(A_T,S_T)].
    \end{equation}
    Let \(T\to \infty\),
    we have
    \[
        v(a,s) \geq \liminf_{T\to\infty}{\E}_{as}\left[
            \int_0^T \me^{-r t}f(A_t, S_t, k_{n,t}) \dd{t}
        \right]
        + \liminf_{T\to\infty} \me^{-r T}{\E}_{as}[v(A_T,S_T)].
    \]
    
    Because \(\int_0^T \me^{-r t}f(A_t, S_t, k_{n,t}) \dd{t}\)
    is nondecreasing in \(T\),
    as the integrand is non-negative,
    we can apply either the monotone convergence theorem
    or Fatou's lemma
    to have
    \begin{align*}
        v(a,s)
        &\geq {\E}_{as}\left[
            \int_0^{\infty} \me^{-r t}f(A_t, S_t, k_{n,t}) \dd{t}
        \right]
        + \liminf_{T\to\infty} \me^{-r T}{\E}_{as}[v(A_T,S_T)] \\
        &=
        v_n(\, a,s \mid k_n,\bm{k}_{\neg n} \,)
        + \liminf_{T\to\infty} \me^{-r T}{\E}_{as}[v(A_T,S_T)] \\
        &\geq 
        v_n(\, a,s \mid k_n, \bm{k}_{\neg n} \,).
    \end{align*}

    Now we repeat the argument by replacing \(k_{n,t}\)
    with \(k^*_{n,t}=k_n^*(A_t)\).
    Inequality~\eqref{eq:hjb-verification} becomes equality,
    and for \(0< T< \infty\) we have
    \begin{align*}
        v(a,s)
        &= {\E}_{as}\left[
            \int_0^T \me^{-r t}f(A_t, S_t, k^*_{n,t}) \dd{t}
        \right]
        +  \me^{-r T}{\E}_{as}[v(A_T,S_T)].
    \end{align*}
    As the first term on the right-hand side is nondecreasing in \(T\),
    the second term must be nonincreasing,
    and hence
    \begin{align*}
        v(a,s)
        &= {\E}_{as}\left[
            \int_0^{\infty} \me^{-r t}f(A_t, S_t, k^*_{n,t}) \dd{t}
        \right]
        +  \lim_{T\to\infty}\me^{-r T}{\E}_{as}[v(A_T,S_T)] \\
        &= v(\, a,s \mid k_n^*,\bm{k}_{\neg n} \,) + \lim_{T\to\infty}\me^{-r T}{\E}_{as}[v(A_T,S_T)].
    \end{align*}
    We finish the proof by showing
    \(\lim_{T\to\infty}\me^{-r T}{\E}_{as}[v(A_T,S_T)] = 0\)
    as follows.

    Because \(A_t \geq 0\),
    we have
    \begin{align*}
        {\E}_{as}[v(A_T,S_T)] &\leq {\E}_{as}[v(0,S_T)] \\
        &= u(0) {\E}_{as}[\me^{S_T}] \\
        &\leq u(0) {\E}_{0s}[\me^{S_T}] \\
        &= u(0) \me^s {\E}_{0s}[\me^{S_T - s}]\\
        &\leq u(0) \me^s {\E}[\me^{M_{NT}}],\\
    \end{align*}
    where
    \(M_T = \max_{0\leq t\leq T}\{\mu t + \sigma B_t\}.\)
    The last inequality comes from the fact that 
    \(S_T - s\leq M_{NT}\) almost surely.
    Indeed,
    given that \(A_0 = 0\),
    the process \(\{S_T - s\}_{T\geq 0}\)
    can be viewed
    as \(\{M_T\}_{T\geq 0}\) with time change,
    in the sense that 
    \(S_T - s = M_{T'}\)
    with \(T'=\int_0^T K_t \dd{t}\).
    Because \(K_t\in[0, N]\),
    we have \(T'\leq NT\)
    and hence \(M_{T'} \leq M_{NT}\).

    Additional derivations based on equation (7.27) in \textcite{Shreve:2004} lead to
    \[
        {\E}[\me^{M_{NT}}]\leq C_1 \me^{(\mu + \sigma^2/2)NT} + C_2
    \]
    for some \(C_1,C_2\geq 0\),
    and therefore we have
    \[
        \me^{-rT}{\E}_{as}[v(A_T,S_T)] \leq 
        C_1 \me^{(-r+(\mu + \sigma^2/2)N) T} + C_2\me^{-rT}.
    \]
    The right-hand side goes to 0 as \(T\to\infty\)
    as long as \((\mu+\sigma^2/2)N < r\),
    which is precisely Assumption~\ref{asm:main-assumption}.
\end{proof}

\begin{lemma}[Necessity]
\label{lem:mpe-necessity}

In any MPE \(\bm{k}\in\mathcal{K}^N\),
for each player \(n \in \{1,\ldots, N\}\),
her payoff function \(u_n(\, \cdot \mid \bm{k} \,)\) satisfies
Conditions~\ref{item:eq-char-1}--\ref{item:eq-char-4} in Lemma~\ref{lem:char-mpe},
and her equilibrium strategy \(k_n(a)\) maximizes
the right-hand side of the HJB equation~\eqref{eq:hjb}
at each continuity point of \(u_n''\).
\end{lemma}
\begin{proof}
    Condition~\ref{item:eq-char-1}:
    In the interior of the stopping region where \(K(a) = 0\),
    it is trivial that the payoff functions are smooth.
    In the region where \(K(a)\) is bounded away from zero,
    players' payoff functions are continuous by standard results,\footnote{
        See, e.g., \textcite{StruloviciSzydlowski:2015},
        footnote 71.
    }
    and once continuously differentiable
    by Lemma~\ref{lem:smooth-payoff}.
    Because the intensity of exploration
    in any MPE is bounded away from zero
    on any compact subset of \([0,\bar{a})\),
    as we argue in Section~\ref{sec:mpe-ppt},
    the only part left to prove 
    is the smooth pasting condition,
    which states that 
    the payoff functions in any MPE must be once continuously differentiable
    at the stopping threshold \(\bar{a}\).
    This is proved in Lemma~\ref{lem:smooth-pasting} below.

    Condition~\ref{item:eq-char-2}:
    In the interior of the stopping region where \(K(a) = 0\),
    it is again trivial that the payoff functions are smooth.
    In the region where \(K(a)\) is bounded away from zero,
    by Lemma~\ref{lem:smooth-payoff},
    players' payoff functions are
    twice continuously differentiable
    at each point at which all players' strategies are continuous.
    Condition~\ref{item:eq-char-2} thus follows from our piecewise Lipschitz-continuity assumption
    on players' strategies.

    Condition~\ref{item:eq-char-3} is a property of any payoff functions
    for any strategy profile with \(K(a)\) bounded away from 0 around \(a=0\).
    See Lemma~\ref{lem:payoff-function-ppt-x}.
      
    Condition~\ref{item:eq-char-4} is proved in Lemma~\ref{lem:hjb} below.
\end{proof}

\begin{lemma}[Smooth Pasting Condition]
    \label{lem:smooth-pasting}
    In any MPE with stopping threshold \(\bar{a} > 0\),
    each player's normalized payoff function
    is continuously differentiable at \(\bar{a}\).
\end{lemma}
\begin{proof}
    Given any MPE \(\bm{k} = (k_1,\ldots, k_N)\),
    consider the region \([0, \bar{a})\)
    in which the intensity of exploration is positive.
    Clearly \(u_n'(\bar{a}-)\),
    the left derivative of
    the equilibrium payoff function of player \(n\) at \(\bar{a}\),
    cannot be positive,
    because otherwise we would have \(u_n(a) < 1\)
    for \(a\) immediately to the left of \(\bar{a}\),
    contradicting the fact that \(k_n\) is a best response.

    Suppose by contradiction that 
    \(u_n\) violates the smooth pasting condition
    at \(\bar{a}\) with \(u_n'(\bar{a}-) < 0\).
    Consider the deviation strategy \(k_\epsilon\)
    in which
    \(k_\epsilon(a) = 1\)
    on \([\bar{a}-\epsilon, \bar{a}+\epsilon)\)
    for some small \(\epsilon > 0\),
    and \(k_\epsilon = k_n\) otherwise,
    with \(u_\epsilon\) denoting the associated payoff function of player \(n\).
    Moreover,
    write \(w_n\coloneqq \left(\ln(u_n)\right)' = u_n'/u_n\)
    and \(w_\epsilon\coloneqq \left(\ln(u_\epsilon)\right)' = u_\epsilon'/u_\epsilon\).
    Note that under this deviation
    the intensity of exploration \(k_\epsilon + \sum_{l\neq n}k_l\)
    is bounded away from zero on \([0, \bar{a}+\epsilon)\)
    so that \(u_\epsilon\)
    is once continuously differentiable in this region,
    which implies \(w_\epsilon\) is continuous on \([0, \bar{a}+\epsilon)\).
    Also, note that because the strategy profile remains unchanged
    on \([0, \bar{a}-\epsilon)\),
    the normal reflection condition \(w_n(0+) = w_\epsilon(0+) = -1\)
    and Feynman-Kac equation~\eqref{eq:feynman-kac-x}
    imply that \(w_n\) and \(w_\epsilon\) coincide on \([0, \bar{a}-\epsilon)\).\footnote{
        \label{fn:int-diff}
        If we let \(w\coloneqq (\ln u)'\) for some payoff function \(u\),
        then Feynman-Kac equation~\eqref{eq:feynman-kac-x}
        can be written in terms of \(w\) as
        \begin{equation}
             1 = \me^{-\int_{a_0}^a w(z)\dd{z}}(1 - k(a)) + K(a)\rho(w'(a) - \theta w(a) + w(a)^2), \label{eq:int-diff}
        \end{equation}
        for any \(a_0\) in the stopping region.
        The normal reflection condition~\eqref{eq:normal-reflection} provides the initial condition \(w(0+) = -1\).
        Also, note that \(w = 0\) in the interior of the stopping region because \(u' = 0\).
    }
    From \(u_n'(\bar{a}-) < 0\) we know that \(w_n(\bar{a}-) < 0\) as well,
    and therefore by the continuity of \(w_\epsilon\)
    we can choose \(\epsilon\) small enough
    so that \(w_\epsilon < 0\) on \([\bar{a} - \epsilon, \bar{a} + \epsilon)\).
    This implies \(u_\epsilon > 1 = u_n\)
    on \([\bar{a}, \bar{a} + \epsilon)\),
    because \(u_\epsilon(a) = \exp(\int_{\bar{a}+\epsilon}^a w_\epsilon(z)\dd{z})\)
    for \(a < \bar{a}+\epsilon\).
    Therefore,
    the deviation strategy \(k_\epsilon\)
    leads to a higher payoff
    on \([\bar{a}, \bar{a} + \epsilon)\)
    than the equilibrium strategy \(k_n\),
    which is a contradiction.
\end{proof}
\begin{lemma}[HJB Equation]
    \label{lem:hjb}
    In any MPE \(\bm{k} = (k_1^*,\ldots, k_N^*)\in\mathcal{K}^N\),
    player \(n\)'s payoff function \(u_n(\, \cdot \mid \bm{k} \,)\)
    satisfies,
    at each continuity point of \(u_n''\), HJB equation~\eqref{eq:hjb}
    with
    \[
        k^*_n(a) \in 
        \argmax_{k_n\in[0,1]}
        k_n\{\beta(a, u_n) - 1\}.
    \]
\end{lemma}
\begin{proof}
    Since for any given \(\bm{k}\),
    player \(n\)'s payoff \(u_n\) satisfies Feynman-Kac Equation~\eqref{eq:feynman-kac-x}
    at each continuity point of \(u_n''\)
    by Lemma~\ref{lem:payoff-function-ppt-x},
    we only need to prove that
    \(k^*_n(a)\) maximizes \(k_n\{\beta(a, u_n) - 1\}\)
    at each continuity point of \(u_n''\).

    If \(a > \bar{a}\) and thus belongs to the stopping region
    in which \(k_n^* = 0\) and \(u_n = 1\),
    we have \(\beta(a, u_n) = 0\) and thus \(k_n^*(a) = 0\) is a maximum.

    If \(a = \bar{a}\) is a continuity point of \(u_n''\),
    then \(\beta(\,\cdot\,, u_n) = 0\) around \(\bar{a}\),
    and hence \(k_n^*(a) = 0\),
    which is required by the right-continuity assumption on the strategies,
    is a maximum.

    We now turn to the case of \(a\in(0, \bar{a})\).
    If \(\beta(\,\cdot\,, u_n) = 1\) in some neighborhood of \(a\),
    then any \(k_n^*(a)\in[0,1]\) is a maximum and there is nothing to prove.
    Next, we prove for the case of \(\beta(a, u_n) > 1\).
    The proof for the case of \(\beta(a, u_n) < 1\) is analogous and thus omitted.

    If \(\beta(a, u_n) > 1\),
    suppose by contradiction that \(k_n^*(a) < 1\).
    Since \(u_n''\) is continuous
    and \(k_n^*\) is right-continuous at \(a\),
    we assume without loss that on \((a, a + \epsilon)\) for some \(a + \epsilon < \bar{a}\),
    \(u_n''\) is continuous,
    \(k_n^*(\cdot) < 1\),
    and \(\beta(\,\cdot\,, u_n) > 1\).

    Consider the deviation strategy \(k_\epsilon\)
    in which \(k_\epsilon = 1\) on \([a, a + \epsilon)\)
    and \(k_\epsilon = k_n^*\) otherwise,
    with \(u_\epsilon\) denoting the associated payoff function of player \(n\).
    Moreover,
    write \(w_n\coloneqq \left(\ln(u_n)\right)' = u_n'/u_n\)
    and \(w_\epsilon\coloneqq \left(\ln(u_\epsilon)\right)' = u_\epsilon'/u_\epsilon\).
    Because the intensity of exploration is bounded away from zero
    on any compact subset of \([0,\bar{a})\),
    Lemma~\ref{lem:smooth-payoff} implies that
    \(w_n\) and \(w_\epsilon\) are continuous
    on \((0, \bar{a})\).
    Also, note that the strategy remains unchanged on \([0, a)\).
    As a consequence,
    the normal reflection condition \(w_n(0+) = w_\epsilon(0+) = -1\)
    and Feynman-Kac equation~\eqref{eq:feynman-kac-x}
    imply that \(w_n\) and \(w_\epsilon\) coincide on \((0, a]\).\footnote{
        See footnote~\ref{fn:int-diff}.
        This claim is not affected by the possible discontinuities of the strategies on \([0, \bar{a})\),
        as the payoff functions are at least once continuously differentiable in this region.
    }

    Rewrite Feynman-Kac equation~\eqref{eq:feynman-kac-x} as
    \((u_n(a) - 1 + k_n(a))/ (K_{\neg n}(a) + k_n(a)) = \beta(a, u_n)\).
    Since \(\beta(a, u_n) > 1\) by assumption,
    we have \(u_n(a) - 1 > K_{\neg n}(a)\).
    This implies
    \((u_n(a) - 1 + k_n)/ (K_{\neg n}(a) + k_n) = \beta(a, u_n) = \rho(u_n''(a) - \theta u'(a))\)
    is decreasing in \(k_n\).
    Because \(u_n, u_\epsilon\) and their derivatives
    are continuous at \(a\),
    and \(k_\epsilon(a+) = 1 > k_n^*(a+)\),
    we have \(u_\epsilon''(a+) < u_n''(a+)\).
    By the definition of \(w_n\) and \(w_\epsilon\),
    we can conclude \(w_\epsilon'(a+) < w_n'(a+)\).
    As \(w_\epsilon(a) = w_n(a)\),
    we can then let \(\epsilon\) be sufficiently small
    so that \(w_\epsilon(a+\epsilon) < w_n(a+\epsilon)\).
    Since both \(w_n\) and \(u_\epsilon\)
    solves the same equation~\eqref{eq:int-diff} in footnote~\ref{fn:int-diff}
    on \((a+\epsilon, \bar{a})\),
    we must have \(w_\epsilon(\bar{a}-) < w_n(\bar{a}-)\).
    This implies \(u_\epsilon'(\bar{a}-) < u_n'(\bar{a}-) = 0\),
    where the equality comes from the smooth pasting condition for equilibrium payoff functions.
    Therefore, the deviation strategy \(k_\epsilon\)
    leads to a payoff
    higher than the equilibrium strategy \(k_n^*\)
    on \((\bar{a} -\delta, \bar{a})\)
    for some \(\delta > 0\),
    which is a contradiction.
\end{proof}

\section{Cooperative Solution}
\label{sec:proofs-coop}
\subsection{Proof of Proposition~\ref{prop:cooperative-solution}}
\begin{proof}
    The cooperative solution can be viewed
    as a corollary of Lemma~\ref{lem:mpe-sufficiency} with \(K_{\neg n} = 0\).
    The closed-form expressions for the stopping threshold and the payoff functions
    follow from the explicit calculation.
\end{proof}

\section{Properties of MPE}
\label{sec:proofs-ppt-mpe}
\subsection{Proof of Proposition~\ref{prop:everyone-explores}}

\begin{proof}
    Suppose to the contrary that there is an MPE where
    player 1 chooses exploitation at all states.
    Let \(\bar{a} > 0\) denote the state at which all exploration stops.
    Then on \((0, \bar{a})\) player 1's payoff function \(u_1\)
    is continuously differentiable and solves
    the free-rider ODE \(u(a) = 1 + K(a)\beta(a, u)\)
    with value matching \(u(\bar{a}) = 1\)
    and smooth pasting condition \(u'(\bar{a}) = 0\).
    It can then be easily verified that
    \(u_1(a) = 1\) is the unique solution,
    which is a contradiction
    because the normal reflection condition~\eqref{eq:normal-reflection}
    is violated.
\end{proof}

\subsection{Proof of Proposition~\ref{prop:encouragement-effct}}

\begin{proof}
    As the individual payoff functions are bounded from below
    by the single-agent payoff function \(U^*_1\),
    it is clear that the stopping threshold \(\bar{a}\)
    in any MPE is weakly larger than the single-agent cutoff \(a^*_1\).
    Suppose by contradiction that
    \(\bar{a} = a^*_1\).
    Assume without loss of generality
    that all players' strategies are continuous
    on \((a^*_1 -\epsilon, a^*_1)\) for some \(\epsilon > 0\),
    so that each player's payoff function
    is twice continuously differentiable in this region.
    Note that for each player \(n\),
    both \(u_n\) and \(U_1^*\)
    satisfy value matching and smooth pasting at \(a^*_1\).
    Then from equation~\eqref{eq:feynman-kac-x},
    we have for each player \(n\),
    \[
        \rho u_{n}''(a^*_1-) = \frac{k_n(a^*_1-)}{K(a^*_1-)}\leq 1
    \]
    and \(\rho U_1^{*\prime\prime}(a^*_1-) = 1\).
    As \(\rho u_{n}''(a^*_1-) < 1\) for at least one player,
    her payoff lower bound \(u_n\geq U_1^*\) is then violated
    at the states immediately to the left of \(\bar{a}\).
\end{proof}

\subsection{Proof of Proposition~\ref{prop:no-cutoff-strategies}}

\begin{proof}
    
    Suppose by contradiction that
    there is an MPE where all players use cutoff strategies.
    Let player 1 be the one who uses the strategy with the largest cutoff \(\bar{a}\).
    Then no other player uses the same cutoff \(\bar{a}\) as player 1
    for the following reason.
    Suppose to the contrary that player 2 uses a strategy with
    the same cutoff \(\bar{a}\),
    then both player 1 and 2 must have a payoff strictly
    greater than 1 and lower than 2
    at the states immediately to the left of \(\bar{a}\),
    as their payoff functions solve the explorer ODE
    \(u(a) = K\beta(a, u)\)
    for some \(K > 1\)
    with the initial condition \(u(\bar{a}) = 1\) and \(u'(\bar{a}) = 0\).
    As a result,
    exploration with full intensity cannot be optimal
    for both players at the states immediately to the left of cutoff \(\bar{a}\)
    according to our characterization of best responses.
    Therefore, player 1 would be the lone explorer
    with \(u_1 > 1\) on \((\bar{a}-\epsilon, \bar{a})\)
    for some \(\epsilon > 0\),
    whereas all other players free-ride in this region.

    Moreover,
    it is easy to see that
    player 1's payoff is weakly lower than the others,
    because the region in which she collects flow payoffs
    is the smallest among all players.
    However, on \((\bar{a}-\epsilon, \bar{a})\)
    the payoff function \(u_n\) of player \(n \neq 1\)
    satisfies the free-rider ODE 
    \(u= 1 + \beta(a, u)\)
    with the same initial conditions as player 1.
    This yields the unique solution \(u_n = 1 < u_1\)
    on \((\bar{a}-\epsilon, \bar{a})\),
    a contradiction.
\end{proof}

\section{Symmetric MPE}
\label{sec:proofs-sym-mpe}
\subsection{Proof of Proposition~\ref{prop:symmetric-mpe}}
From Lemma~\ref{lem:char-mpe},
it is not difficult to verify that
the strategy profile
in Corollary~\ref{cor:symmetric-mpe}
constitutes an equilibrium,
and that our proposition properly summarizes
the equilibrium payoff function
in that corollary.

Uniqueness follows directly from symmetry
and Lemma~\ref{lem:char-mpe}.
One can check
by explicit calculations
that,
under Assumption~\ref{asm:main-assumption},
the equilibrium in Corollary~\ref{cor:symmetric-mpe}
is the only symmetric \(s\)-invariant strategy profile
such that the associated common payoff function
satisfies Conditions~\ref{item:eq-char-1}--\ref{item:eq-char-4} in Lemma~\ref{lem:char-mpe}.

The comparison between the stopping threshold
and the cooperative cutoff
follows from Lemma~\ref{lem:welfare-comparison}.

\section{Comparative Statics}
\label{sec:proofs-cs}
\begin{lemma}[Welfare Comparison]
    \label{lem:welfare-comparison}
    Consider normalized payoff functions
    \(u_j:\R_{+}\to[1, +\infty)\) for \(j\in\{l,h\}\),
    with stopping thresholds \(\bar{a}_j\coloneqq \sup\{\,a\geq 0 \mid u_j(a) > 1\,\} < +\infty\).
    Suppose for each \(j\in\{l,h\}\),
    \(u_j\) satisfies the following conditions:
    \begin{enumerate}[label=(\roman*)]
        \item \(C^1\) on \((0, \bar{a}_j)\)
        with \(u_l'(\bar{a}_l-) = u_h'(\bar{a}_h-)\leq 0\);
        \label{item:lem-welfare-comparison-1}
        \item piecewise \(C^2\) on \(\R_{++}\);
        \label{item:lem-welfare-comparison-2}
        \item 
        there exists some function \(g_j:(1,+\infty)\times\R_{-}\to \R_{++}\)
        with
        \(g_j(z_1,z_2)/z_2\) nondecreasing in \(z_1\) and nonincreasing in \(z_2\)
        such that
        \begin{enumerate}
            \item \(u_j''(a) = g_j(u_j(a), u_j'(a))\)
        for all \(a\in(0, \bar{a}_j)\) at which \(u_j''(a)\) is continuous;
            \item  \label{item:lem-welfare-comparison-b}
            \(u_j'(a) + g_j(u_j(a), u_j'(a)) > 0\)
        for all \(a\in(0, \bar{a}_j)\).
        \end{enumerate}
        \label{item:lem-welfare-comparison-3}
    \end{enumerate}

    If \(g_l \geq g_h\),
    then \(\bar{a}_l \leq \bar{a}_h\) and \(u_l \leq u_h\).
    Additionally,
    if for some \(\breve{u} \in (1, u_h(0)]\)
    we have
    \(g_l = g_h\)
    on \((\breve{u}, u_h(0))\times\R_{-}\),
    and for each \(z_2 \leq 0\), 
    at least one of the following two conditions holds:
    \begin{enumerate}[label=(\roman*), resume]
        \item \label{item:lem-welfare-comparison-4}
        \(g_l(\breve{u}-, z_2) > g_h(\breve{u}-, z_2)\);
        \item \label{item:lem-welfare-comparison-5}
        \(\pdv{g_l(\breve{u}-, z_2)}{z_1} < \pdv{g_h(\breve{u}-, z_2)}{z_1}\)
        and \(\pdv{g_l(\breve{u}-, z_2)}{z_2} \geq \pdv{g_h(\breve{u}-, z_2)}{z_2}\);
    \end{enumerate}
    then \(\bar{a}_l < \bar{a}_h\), and \(u_l < u_h\) on \([0, \bar{a}_h)\).
\end{lemma}

\begin{remark}
    Almost all payoff functions in this paper
    fulfill Conditions~\ref{item:lem-welfare-comparison-1}--\ref{item:lem-welfare-comparison-3} above
    under Assumption~\ref{asm:main-assumption}
    with value matching and smooth pasting at \(\bar{a}\).
    These include
    \begin{itemize}
        \item the cooperative solution:
        \(g(z_1, z_2) = \frac{z_1}{N\rho} +\theta z_2\);
        \item the symmetric MPE:
        \(g(z_1, z_2) = \max\{z_1/N, 1\}/\rho +\theta z_2\);
        \item the average payoff in the asymmetric MPE:
        \[
            g(z_1, z_2) = \max\{z_1/N, \min\{1, z_1 - (1-1/N)\}\}/\rho +\theta z_2;
        \]
    \end{itemize}
\end{remark}

\begin{proof}[Proof of Lemma~\ref{lem:welfare-comparison}] 
    From \(u_j''(a) = g_j(u_j(a), u_j'(a)) > 0\)
    and \(u_j'(\bar{a}_j-)\leq 0\),
    we know \(u_j'\) is negative on \((0, \bar{a}_j)\).
    Therefore, \(u_j^{-1}(u)\), the inverse of \(u_j\) at \(u\),
    is uniquely defined for \(u\in[1, u_j(0)]\).\footnote{
        We define \(u_j^{-1}(1) \coloneq \bar{a}_j\).
    }
    To simplify notations,
    we denote the derivatives (one-side if necessary)
    of \(u_j\) at level \(u\in [1, u_j(0)]\)
    by 
    \(\tilde{u}_j'(u)\) and \(\tilde{u}_j''(u)\),
    where
    \(\tilde{u}_j' = u_j'\circ u_j^{-1}\)
    and \(\tilde{u}_j'' = u_j''\circ u_j^{-1}\).

    We first show that
    for all \(u\in[1, \min\{u_l(0),u_h(0)\}]\),
    we have \(\tilde{u}_l'(u)\leq \tilde{u}_h'(u)\).
    Suppose by contradiction that
    \(\tilde{u}_l'(\hat{u}) > \tilde{u}_h'(\hat{u})\) for some \(1 \leq \hat{u}\leq \min\{u_l(0),u_h(0)\}\).
    Let \(\check{u}\coloneqq \sup\{\,u\leq\hat{u} \mid \tilde{u}_l'(u)\leq \tilde{u}_h'(u)\,\}\).
    Obviously we have \(1 \leq \check{u} < \hat{u}\)
    since \(\tilde{u}_l'(1) = \tilde{u}_h'(1)\),
    and by  the continuity of \(\tilde{u}_j'\)
    we have \(\tilde{u}_l'(\check{u}) = \tilde{u}_h'(\check{u}) < 0\).
    We then compare their derivatives on \((\check{u},\check{u}+\epsilon)\)
    for some small enough \(\epsilon > 0\) and have
    \[
        \dv{\tilde{u}_l'(u)}{u} 
        = \frac{\tilde{u}_l''(u)}{\tilde{u}_l'(u)}
        = \frac{g_l(u, \tilde{u}_l'(u))}{\tilde{u}_l'(u)}
        \leq \frac{g_l(u, \tilde{u}_h'(u))}{\tilde{u}_h'(u)}
        \leq \frac{g_h(u, \tilde{u}_h'(u))}{\tilde{u}_h'(u)}
        = \frac{\tilde{u}_h''(u)}{\tilde{u}_h'(u)}
        = \dv{\tilde{u}_h'(u)}{u},
    \]
    where the first inequality comes from the monotonicity of \(g_l(z_1, z_2)/z_2\)
    in \(z_2\),
    and the second inequality comes from \(\tilde{u}_h'(u) < 0\)
    and our assumption that \(g_l \geq g_h > 0\).
    This, together with \(\tilde{u}_l'(\check{u}) = \tilde{u}_h'(\check{u})\),
    implies \(\tilde{u}_l'(u) \leq \tilde{u}_h'(u)\) on \((\check{u},\check{u}+\epsilon)\),
    contradicting the definition of \(\check{u}\).
    
    Next we show that \(u_l(0) \leq u_h(0)\).    
    Suppose by contradiction that \(u_l(0) > u_h(0)\).
    Then the normal reflection conditions \(u_j(0) = - u_j'(0)\)
    and Condition (\ref{item:lem-welfare-comparison-b}) in the lemma
    imply the following strict inequality
    \begin{align*}
        u_l(0) = -u_l'(0) = -\tilde{u}_l'(u_l(0))
        &= -\tilde{u}_l'(u_h(0)) - \int_{u_h(0)}^{u_l(0)} \tilde{u}_l''(u)/\tilde{u}_l'(u)\dd{u} \\
        &= -\tilde{u}_l'(u_h(0)) - \int_{u_h(0)}^{u_l(0)} g_l(u, \tilde{u}_l'(u))/\tilde{u}_l'(u)\dd{u} \\
        &\geq -\tilde{u}_h'(u_h(0)) - \int_{u_h(0)}^{u_l(0)} g_l(u, \tilde{u}_l'(u))/\tilde{u}_l'(u)\dd{u} \\
        &> -\tilde{u}_h'(u_h(0)) + \int_{u_h(0)}^{u_l(0)} 1\dd{u} \\
        &= -u_h'(0) + u_l(0) - u_h(0) = u_l(0),
    \end{align*}
    which is a contradiction.

    Since we have
    \(\tilde{u}_l'(u) \leq \tilde{u}_h'(u)\leq 0\)
    for all \(u\in [ 1, u_l(0)]\) with \(u_l(0) \leq u_h(0)\),
    it is obvious that \(\bar{a}_l\leq \bar{a}_h\)
    and \(u_l \leq u_h\).

    To prove the strict inequalities,
    suppose in addition to \(g_l\geq g_h\)
    we also have 
    \(g_l = g_h\)
    on \((\breve{u}, u_h(0))\times\R_{-}\)
    for some \(\breve{u}\in(1, u_h(0)]\).
    Moreover,
    assume that for each \(z_2\leq 0\),
    we have either~\ref{item:lem-welfare-comparison-4}
    \(g_l(\breve{u}-, z_2) > g_h(\breve{u}-, z_2)\),
    or we have~\ref{item:lem-welfare-comparison-5}
    \(\pdv{g_l(\breve{u}-, z_2)}{z_1} < \pdv{g_h(\breve{u}-, z_2)}{z_1}\)
    and \(\pdv{g_l(\breve{u}-, z_2)}{z_2} \geq \pdv{g_h(\breve{u}-, z_2)}{z_2}\).
    We next show that \(u_l(0) < u_h(0)\),
    which together with the weak inequalities proved above
    yields the strict inequalities in the Lemma.
    
    Suppose by contradiction that
    \(u_l(0) = u_h(0)\).
    Note that either of~\ref{item:lem-welfare-comparison-4} or~\ref{item:lem-welfare-comparison-5}
    implies \(g_l(z_1, z_2) > g_h(z_1, z_2)\)
    for \(z_1\) immediately below \(\breve{u}\),
    which together with \(\tilde{u}_l'(\breve{u}) \leq \tilde{u}_h'(\breve{u})\) proved above
    implies that \(\tilde{u}_l' < \tilde{u}_h' < 0\) on \((\breve{u}-\eta, \breve{u})\)
    for some \(\eta > 0\).
    Moreover,
    by the normal reflection conditions,
    our assumption \(u_l(0) = u_h(0)\) implies \(u_l'(0) = u_h'(0)\).
    Since \(g_l = g_h\)
    on \((\breve{u}, u_h(0))\times\R_{-}\),
    and the initial conditions for 
    \(u_j'' = g_j(u_j, u_j')\) are the same for both \(j=l,h\),
    we must have \(\tilde{u}_l'(\breve{u}) = \tilde{u}_h'(\breve{u})\eqqcolon \bar{z}_2 < 0\).
    We can then compare the left derivatives
    of \(\tilde{u}_j'\) at \(\breve{u}\) to have
    \[
        \dv{\tilde{u}_l'(\breve{u}-)}{u} 
        = \frac{\tilde{u}_l''(\breve{u}-)}{\tilde{u}_l'(\breve{u})}
        = \frac{g_l(\breve{u}-, \bar{z}_2)}{\bar{z}_2}
        \leq \frac{g_h(\breve{u}-, \bar{z}_2)}{\bar{z}_2}
        = \frac{\tilde{u}_h''(\breve{u}-)}{\tilde{u}_h'(\breve{u})}
        = \dv{\tilde{u}_h'(\breve{u}-)}{u}.
    \]
    If \(g_l(\breve{u}-, \bar{z}_2) > g_h(\breve{u}-, \bar{z}_2)\),
    then this inequality is strict,
    which together with \(\tilde{u}_l'(\breve{u}) = \tilde{u}_h'(\breve{u})\),
    contradicts the fact that \(\tilde{u}_l' < \tilde{u}_h'\)
    on \((\breve{u}-\eta, \breve{u})\).
    Otherwise,
    it must be
    \(g_l(\breve{u}-, \bar{z}_2) = g_h(\breve{u}-, \bar{z}_2)\)
    and hence    
    \(\dv{\tilde{u}_l'(\breve{u}-)}{u} = \dv{\tilde{u}_h'(\breve{u}-)}{u} < 0\).
    Then from their second-order derivatives
    \begin{align*}
        \dv[2]{\tilde{u}_j'(u)}{u} 
        &= \frac{1}{\tilde{u}_j'(u)}\dv{g_j(u, \tilde{u}_j'(u))}{u}
        - \frac{g_j(u, \tilde{u}_j'(u))}{(\tilde{u}_j'(u))^2}\dv{\tilde{u}_j'(u)}{u} \\
        &= \frac{1}{\tilde{u}_j'(u)}
        \left(
            \pdv{g_j(u, \tilde{u}_j'(u))}{z_1} + \pdv{g_j(u, \tilde{u}_j'(u))}{z_2} \dv{\tilde{u}_j'(u)}{u}  
        \right)
        - \frac{g_j(u, \tilde{u}_j'(u))}{(\tilde{u}_j'(u))^2}\dv{\tilde{u}_j'(u)}{u},
    \end{align*}
    we can conclude from
    \(\pdv{g_l(\breve{u}-, \bar{z}_2)}{z_1} < \pdv{g_h(\breve{u}-, \bar{z}_2)}{z_1}\)
    and \(\pdv{g_l(\breve{u}-, \bar{z}_2)}{z_2} \geq \pdv{g_h(\breve{u}-, \bar{z}_2)}{z_2}\)
    that
    \(\dv[2]{\tilde{u}_l'(\breve{u}-)}{u} > \dv[2]{\tilde{u}_h'(\breve{u}-)}{u}\).
    This, together with
    \(\tilde{u}_l'(\breve{u}) = \tilde{u}_h'(\breve{u})\)
    and
    \(\dv{\tilde{u}_l'(\breve{u}-)}{u} = \dv{\tilde{u}_h'(\breve{u}-)}{u}\),
    contradicts the fact that \(\tilde{u}_l' < \tilde{u}_h'\)
    on \((\breve{u}-\eta , \breve{u})\).
\end{proof}

\subsection{Proof of Corollary~\ref{cor:cs-coop}}
\begin{proof}
    Here we prove the limit results only.
    Other results can be easily derived from explicit calculations.
    Let \(\widehat{U}_N = 1 + \exp(-\lambda a)/(\lambda - 1)\) with \(\lambda \coloneqq 1/(N\rho) - \theta\),
    and write \(\hat{w}_N = (\ln\widehat{U}_N)'\) and \(w^*_N = (\ln U^*_N)'\).
    It is not difficult to verify that
    \(\hat{w}_N' = -\left(\frac{1}{N\rho} - \theta\right) \hat{w}_N - \hat{w}_N^2\)
    on \(\R_{++}\);
    \(w_N^{*\prime} = \frac{1}{N\rho} + \theta w^*_N - {w^*_N}^2\)
    on \((0, a^*_N)\)
    and \(w^*_N = 0\)
    on \([a^*_N, +\infty)\).

    Suppose that \(\theta \leq -1\).
    Then on \((0, a^*_N)\),
    both ODEs above converge to
    \(w' - \theta w + w^2 = 0\)
    as \(N\to +\infty\).
    Because the normal reflection condition~\eqref{eq:normal-reflection}
    implies \(w^*_N(0) = -1 = \hat{w}_N(0)\),
    we must have
    \(\lim_{N\to\infty} w^*_N = \lim_{N\to\infty}\hat{w}_N\)
    on \((0, a^*_N)\).
    However, note that \(\lim_{N\to\infty} \hat{w}_N(a) < 0\) for all \(a\geq 0\).
    Therefore, we must have \(a^*_N\to +\infty\)
    by the continuity of \(w^*_N\) on \(\R_{+}\).
    This, together with the convergence of \(w^*_N\) and \(\hat{w}_N\),
    implies
    \(\lim_{N\to\infty} U^*_N(a) = \lim_{N\to\infty}\widehat{U}_N(a)\)
    for each \(a\geq 0\).
    In particular for \(\theta = -1\),
    we can see that \(\widehat{U}_N(a)\to+\infty\) as \(N\to+\infty\),
    and hence \(U^*_N(a) \to +\infty\) for each \(a\geq 0\).
    Then the results for the case of \(\theta > -1\)
    follow from the monotonicity of \(U^*_N(a)\) in \(\theta\),
    which can be easily verified by Lemma~\ref{lem:welfare-comparison}.
\end{proof}

\subsection{Proof of Corollary~\ref{cor:cs-n}}
\begin{proof}

    For the binding case,
    the comparative statics of \(U^\dagger_N\)
    and \(\tilde{a}_N\)
    with respect to \(N\)
    are immediate from Lemma~\ref{lem:welfare-comparison}.

    For the non-binding case,
    Lemma~\ref{lem:welfare-comparison} implies that
    both \(U^\dagger_N\) and \(\tilde{a}_N\) are constant over \(N\),
    as \(g(z_1,z_2) = 1/\rho +\theta z_2\) does not depend on \(N\).
    Then it is obvious that \(k_N(x) = (U^\dagger_N(a) - 1)/ (N-1)\) is weakly decreasing in \(N\) for each \(a\geq 0\).
\end{proof}

\subsection{Proof of Corollary~\ref{cor:asymp-n}}
For the sake of notational convenience,
we are going to prove
the results in terms of \(\rho  = \sigma^2/(2r) > \hat{\rho}\)
for some \(\hat{\rho}\),
which is equivalent to \(r < \hat{r} = \sigma^2/(2\hat{\rho})\).
We let
\(\hat{\rho}
= (\theta - \ln(1+\theta))/\theta^2\)
if \(\theta > -1\).\footnote{If \(\theta=0\) we set \(\hat{\rho} = 1/2\), the limit of the right-hand side as \(\theta\to 0\).}

To show the statement for the case that
\(\rho > \hat{\rho}\) and \(\theta > -1\),
we first show in Lemma~\ref{lem:finite-gap} below that
\(\lvert a^*_N - a^\dagger_N\rvert \to \Delta\)
for some \(\Delta < +\infty\)
as \(N\to\widehat{N}\coloneqq 1/(\rho(1+\theta))\).
Because we know from Corollary~\ref{cor:cs-coop} that
\(a^*_N\to +\infty\),
we then conclude that \(a^\dagger_N\to +\infty\),
and hence \(\tilde{a}_N\to +\infty\) as well.

After showing that
\(\lim_{N\to\widehat{N}} \norm{U^*_N/U^\dagger_N}_\infty < +\infty\)
in Lemma~\ref{lem:bounded-payoffs},
we can then conclude \(U^\dagger_N\to +\infty\)
from the fact that \(U^*_N\to +\infty\)
from Corollary~\ref{cor:cs-coop}.

The opposite case
that \(\rho \leq \hat{\rho}\) or \(\theta \leq -1\)
follows from Lemma~\ref{lem:asymp-n} below.

\begin{lemma}
    \label{lem:finite-gap}
    If \(\rho > \hat{\rho}\),
    we have \(\lvert a^*_N - a^\dagger_N\rvert \to\Delta\in(0,+\infty)\) as \(N\to\widehat{N}\).
\end{lemma}

\begin{proof}

    Note that when \(\rho > \hat{\rho}\),
    the expression of \(\tilde{a}_N\)
    for the non-binding case in Corollary~\ref{cor:symmetric-mpe}
    cannot be applied.
    Therefore, the symmetric MPE must belong to the binding case for all \(N\in(1,\widehat{N})\).
    Moreover,
    for \(\rho > \hat{\rho}\),
    we can verify that
    \(\iota_N\) in Corollary~\ref{cor:symmetric-mpe}
    converges to some \(\iota_{\widehat{N}}\in(0,1)\).
    Then from explicit calculations,
    we know that
    \(\gamma_2\to  1 + \theta > 0\)
    and \(\gamma_1 \to -1\) as \(N\to\widehat{N}\),
    and therefore we have
    \begin{align*}
        \Delta &\coloneqq \lim_{N\to\widehat{N}}
        \lvert a^*_N - a^\dagger_N\rvert
        = \lim_{N\to\widehat{N}} \frac{1}{\gamma_2 - \gamma_1}\ln\left(\frac{1  +\iota/\gamma_2}{1+\iota/\gamma_1}\right) \\
        &= \frac{1}{2+\theta}\ln\left(\frac{1+\iota_{\widehat{N}}/(1+\theta)}{1-\iota_{\widehat{N}}}\right)
        \in(0,+\infty).
    \end{align*}
\end{proof}

\begin{lemma}
    \label{lem:bounded-payoffs}
    If \(\rho > \hat{\rho}\),
    we have \(\lim \norm{U^*_N/U^\dagger_N}_\infty < +\infty\)
    as \(N\to\widehat{N}\).
\end{lemma}
\begin{proof}

    We first show that
    \(\norm{U^*_N/U^\dagger_N}_\infty = U^*_N(0)/U^\dagger_N(0)\)
    by showing
    \(U^*_N(a)/U^\dagger_N(a)\) is nonincreasing in \(a\).

    Write \(w^*_N = (\ln U^*_N)'\) and \(w^\dagger_N = (\ln U^\dagger_N)'\).
    It is not difficult to verify that
    \begin{align*}
        w_N^{*\prime} &= \frac{1}{N\rho} + \theta w^*_N - {w^*_N}^2,\quad\text{ on }(0, a^*_N),\\
        \text{and}\quad
        w_N^{\dagger\prime} &= \begin{cases}
            \frac{1}{N\rho} + \theta w^\dagger_N - {w^\dagger_N}^2, & \text{ on }(0, a^\dagger_N), \\
            \frac{1}{U^\dagger_N\rho} + \theta w^\dagger_N - {w^\dagger_N}^2, & \text{ on }(a^\dagger_N, \tilde{a}_N).
        \end{cases}
    \end{align*}
    Because, \(U^\dagger_N < N\) on \((a^\dagger_N, \tilde{a}_N)\),
    we have \(w_N^{\dagger\prime} \geq w_N^{*\prime}\).
    This, together with the normal reflection condition \(w^*_N(0) = w^\dagger_N(0) = -1\),
    implies
    \(w^*_N \leq w^\dagger_N\).
    Therefore,
    \(w^*_N(a)-w^\dagger_N(a) = \left(\ln(U^*_N(a)/U^\dagger_N(a))\right)'\leq 0\),
    which implies
    \(\ln(U^*_N(a)/U^\dagger_N(a))\) is nonincreasing in \(a\)
    and hence so is \(U^*_N(a)/U^\dagger_N(a)\).
    Therefore, we have
    \(\norm{U^*_N/U^\dagger_N}_\infty = U^*_N(0)/U^\dagger_N(0)\).

    Moreover,
    because \(w^*_N(0) = w^\dagger_N(0) = -1\)
    and \(w_N^{*\prime} = w_N^{\dagger\prime}\) on \((0, a^\dagger_N)\),
    we have \(w^*_N = w^\dagger_N\) on \((0, a^\dagger_N)\)
    and hence \(U_N^*/U^\dagger_N\) is constant on \((0, a^\dagger_N)\).
    Therefore, we can write
    \[
        \frac{U^*_N(0)}{U^\dagger_N(0)} = \frac{U^*_N(a^\dagger_N)}{U^\dagger_N(a^\dagger_N)}
        = \frac{1}{N}\frac{1}{\gamma_2 - \gamma_1}
        \left(
            \gamma_2 \me^{-\gamma_1(a^*_N - a^\dagger_N)}
            -\gamma_1 \me^{-\gamma_2(a^*_N - a^\dagger_N)}
        \right).
    \]
    Since we know that
    \(\gamma_2\to 1+\theta > 0\) and \(\gamma_1 \to -1\),
    we have
    \[
        \norm{\frac{U^*_N}{U^\dagger_N}}_\infty
        =\frac{U^*_N(0)}{U^\dagger_N(0)} \to
        \frac{1}{\widehat{N}(2+\theta)}\left(
            (1+\theta) \me^{\Delta} + \me^{-(1+\theta)\Delta}
        \right) < +\infty.
    \]
\end{proof}

\begin{lemma}
    \label{lem:asymp-n}

    If either \(\rho \leq \hat{\rho}\) or \(\theta \leq -1\),
    then there exists \(\check{N} > 1\)
    such that \(a^\dagger_N = 0\) and \(U^\dagger_N = U^\dagger_{\check{N}}  < +\infty\)
    for all \(N \geq \check{N}\).

\end{lemma}
Here we do not impose Assumption~\ref{asm:main-assumption} in Lemma~\ref{lem:asymp-n}.
Therefore, this lemma states that
when \(N\) reaches \(\check{N}\) as increased from 1,
the symmetric equilibrium falls into the non-binding case
and continues to be an equilibrium for all \(N > \check{N}\),
even when Assumption~\ref{asm:main-assumption} is violated for large \(N\)
in the case of \(\theta > -1\).
As a consequence,
we have \(\tilde{a}_N = \tilde{a}_{\check{N}}\)
for all \(N\geq \check{N}\),
and thus the stopping threshold \(\tilde{a}_N\)
is bounded from above as \(N\to +\infty\).\footnote{
    More precisely,
    when Assumption~\ref{asm:main-assumption}
    is violated and the uniqueness of the symmetric MPE is not guaranteed,
    here we mean there exists a sequence of symmetric MPE indexed by \(N\),
    with \(\tilde{a}_N\) bounded from above as \(N\to +\infty\).
}  
This implies
\(\lvert U^*_N(a) - U^\dagger_N(a)\rvert\)
is bounded away from zero in the limit,
because for large enough \(N\)
we have that \(U^\dagger_N(a)\) is constant over \(N\),
and that \(U^*_N(a)\) is increasing in \(N\)
for each \(a\geq 0\).

Next, we prove the above lemma.

\begin{proof}[Proof of Lemma~\ref{lem:asymp-n}]
    
    Suppose either \(\rho \leq \hat{\rho}\) or \(\theta \leq -1\).
    We construct \(U^\dagger_{\check{N}}\)
    according to the closed-form expression
    of the equilibrium payoff function for the non-binding case
    in Corollary~\ref{cor:symmetric-mpe}.
    This is possible only when
    either \(\rho \leq \hat{\rho}\) or \(\theta \leq -1\),
    as otherwise the expression for \(\tilde{a}_N\)
    in that corollary is not well-defined.
    Let \(\check{N} = U^\dagger_{\check{N}}(0)\),
    and for all \(N\geq\check{N}\),
    let \(k^\dagger_N = (U^\dagger_{\check{N}} - 1)/(N-1)\).

    We now verify that the strategy profile \(\bm{k}^\dagger_N\)
    with each player playing \(k^\dagger_N\)
    constitutes a symmetric MPE 
    with non-binding resource constraints
    in the \(N\)-player exploration game
    for all \(N\geq\check{N}\),
    with \(U^\dagger_{\check{N}}\) being the associated payoff function.
    Note that we cannot apply Lemma~\ref{lem:char-mpe} directly
    because it relies on Assumption~\ref{asm:main-assumption}.
    Nevertheless,
    from the proof of Lemma~\ref{lem:mpe-sufficiency},
    we know that \(U^\dagger_{\check{N}}\)
    is an upper bound on player \(n\)'s achievable payoffs
    against \(\bm{k}^\dagger_{\neg n}\)
    (this fact does not rely on Assumption~\ref{asm:main-assumption}).
    Moreover,
    \(U^\dagger_{\check{N}}\)
    is indeed the payoff function associated with \(\bm{k}^\dagger_N\),
    since
    \(U^\dagger_{\check{N}}\)
    is the only function that satisfies
    both of the properties in Lemma~\ref{lem:payoff-function-ppt-x}.
    As a result,
    the upper bound on the payoff functions
    when the player plays against \(\bm{k}^\dagger_{\neg n}\)
    is achieved by \(k^\dagger_N\),
    and therefore \(\bm{k}^\dagger_N\) is a symmetric MPE.
\end{proof}

\section{Asymmetric MPE}
\label{sec:proofs-asym-mpe}

\subsection{Proof of Proposition~\ref{prop:asymmetric-mpe}}
\subsubsection*{Step 1: Construction of the average payoff function.}

Let \(\tilde{u}:\R_{-}\to\R\) be 
once continuously differentiable and solves
\begin{equation}
    \label{eq:ode-avg-payoff-asym}
    \max\{u(a)/N, \min\{1, u(a) - (1 - 1/N)\}\} = \beta(a, u),
\end{equation}
with the initial conditions \(u(0) = 1\) and \(u'(0) = 0\).\footnote{
    It is straightforward to verify that
    \(\tilde{u}\) is strictly decreasing.
}
Let \(a^\flat_N > 0\)
be such that
\(\tilde{u}(-a^\flat_N) + \tilde{u}'(-a^\flat_N) = 0\).\footnote{
    It can be shown that
    under Assumption~\ref{asm:main-assumption}
    there exists a unique \(0 < a^\flat_N < +\infty\).
}

Now, we let the average payoff function \(\bar{u}(a) = \tilde{u}(a - a^\flat_N)\)
for \(a\leq a^\flat_N\)
and \(\bar{u}(a) = 1\) for \(a > a^\flat_N\).
We can check that \(\bar{u}\) satisfies
value matching \(\bar{u}(a^\flat_N) = 1\),
smooth pasting \(\bar{u}'(a^\flat_N) = 0\),
and the normal reflection condition \(\bar{u}(0) + \bar{u}'(0) = 0\).
Let \(a^\sharp_N = \bar{u}^{-1}(2-1/N)\)
and \(a^\ddag_N = \bar{u}^{-1}(N)\)
where \(\bar{u}^{-1}(u) = \inf\{\,a \geq 0 \mid \bar{u}(a)\leq u\,\}\).
Under such construction we have
\(0 \leq a^\ddag_N \leq a^\sharp_N < a^\flat_N\)
and
\(\bar{u}:\R_{+}\to\R\)
is continuously differentiable on \(\R_{++}\)
and satisfies,
on \((a^\flat_N, +\infty)\), \(\bar{u} = 1\);
on \((a^\sharp_N, a^\flat_N)\), solves \(u = 1 - 1/N + \beta(a, u)\);
on \((a^\ddag_N, a^\sharp_N)\), solves \(1 = \beta(a, u)\);
on \((0, a^\ddag_N)\), solves \(u = N \beta(a, u)\).

\subsubsection*{Step 2: Construction of the players' payoff functions and strategies.}

For each player \(n\),
on \([a^\flat_N, +\infty)\), let \(k_n(a) = 0\) and \(u_n = \bar{u} = 1\);
on \([0, a^\sharp_N)\), if not empty,
let \(k_n(a) = \min\{1, (\bar{u}(a) - 1)/(N-1)\}\)
and \(u_n = \bar{u}\).

Next,
consider any partition
\(a^\sharp_N = a_1 < a_2 < \cdots < a_m < a_{m+1} = a^\flat_N\)
of interval \([a^\sharp_N, a^\flat_N)\).
For each subinterval \([a_j, a_{j+1})\)
in the partition \(\{a_j\}_{j=1}^{m+1}\),
we use Algorithm~\ref{alg:payoff-asym-mpe}
to construct payoff function \(u_n\)
and strategy \(k_n\)
for each player \(n\).

For each subinterval \([a_j, a_{j+1})\),
Algorithm~\ref{alg:payoff-asym-mpe} calls procedure \(\Call{ActionAssignment}{}\),
which first calls function \(\Call{Split}{}\)
to split \([a_j, a_{j+1})\) into three subintervals
according to Lemma~\ref{lem:split} below,
and lets the player with the lowest index available
free-ride on the subinterval \([a_{M-}, a_{M+})\) in the middle,
and explore on the rest two subintervals at both ends.
In such a way, the strategy \(k_n\)
and payoff function \(u_n\) of this player are defined on \([a_j, a_{j+1})\),
and she is then labeled as unavailable.
Then \(\Call{ActionAssignment}{}\)
is called recursively on these three subintervals,
preserving the total intensity \(K=1\) by
allocating intensity \(0\) on the subintervals at both ends,
and intensity \(1\) on \([a_{M-}, a_{M+})\),
with \(\bar{u}\) on these subintervals
being replaced by \(\bar{u}_{\neg n}\),
which is the average payoff function
among the rest of the available players.

Lemma~\ref{lem:split} ensures that function \(\Call{Split}{}\)
partitions any interval \([a_L,a_R]\)
into three subintervals in a unique way
such that \(u_n\),
and therefore also \(\bar{u}_{\neg n}\),
have the same values and derivatives as \(\bar{u}\)
at the end points \(a_L\) and \(a_R\).

The termination of Algorithm~\ref{alg:payoff-asym-mpe} is trivial
because each time \(\Call{ActionAssignment}{}\) is called,
the strategy of one of the players is assigned
and she is thereafter removed from the set of the available players.
In the case where the allocation of strategies is unambiguous---that is,
either the task of exploration has already been assigned
or is to be assigned to the last available player---splitting the interval is unnecessary,
and the corresponding strategies are set at once for all the available players.

When Algorithm~\ref{alg:payoff-asym-mpe} terminates,
the strategies and
the constructed payoff functions of each player
are defined  on \([a^\sharp_N, a^\flat_N)\).
Also, note that
\(\bar{u}\) from Step 1 is indeed the average of the constructed
payoffs of all the players,
and the input requirement for \(\bar{u}\)
in procedure \textsc{Split}
is satisfied in each call to the procedure.
These conditions are maintained by line~\ref{algline:avg-others}
during each call of \(\Call{ActionAssignment}{}\).
Lastly,
\(u_n\) is once continuously differentiable on \(\R_{++}\) for each \(n\),
since no ``kink'' is created in Algorithm~\ref{alg:payoff-asym-mpe}.

\begin{algorithm}
    \caption{Payoff construction}
    \label{alg:payoff-asym-mpe}
    \begin{algorithmic}[1]
    
    \Require $a^\sharp_N = a_1 < \ldots < a_{m+1} = a^\flat_N$, $\bar{u}$ from Step 1, and a finite set of players $\mathcal{N} = \{1,\ldots, N\}$
    \Ensure equilibrium strategy profile $\{k_n\}_{n=1}^{N}$ is defined on \([a_1, a_{m+1}]\),
    and $\{u_n\}_{n=1}^{N}$ is the set  payoff functions corresponding to strategy profile $\{k_n\}_{n=1}^{N}$
    \Statex
    
    \ForAll{$j\in\{1,\ldots, m\}$}
            \State \Call{ActionAssignment}{$\mathcal{N}$, $1$, $a_j$, $a_{j+1}$, $\bar{u}$} \label{algline:initial-call-allocate}
    \EndFor
    \Statex
    
    \Function {Split}{$\abs{\mathcal{I}}$, $a_L$, $a_R$, $\bar{u}$}
    \Comment {input and output according to Lemma~\ref{lem:split}}
    
    \Require $\abs{\mathcal{I}}$: the number of available players
    \Require $\bar{u}$: solves $u = 1 - 1/\abs{\mathcal{I}} + \rho (u '' - \theta u')$ on \((a_L, a_R)\)
    \Ensure $\check{u}\in C^0([a_L, a_R])\cap C^1((a_L,a_R))$ solves $u = \rho (u '' - \theta u')$
    on $(a_L, a_{M-})\cup(a_{M+}, a_R)$,
    and $u = 1 + \rho(u '' - \theta u')$
    on $(a_{M-}, a_{M+})$,
    with the same values and derivatives as \(\bar{u}\) at \(a_L\) and \(a_R\)
    \Statex
    
    \State \Return ($a_{M-}$, $a_{M+}$, $\check{u}$)
    \EndFunction
    \Statex
    
    \Procedure {ActionAssignment}{$\mathcal{I}$, $\kappa$, $a_L$, $a_R$, $\bar{u}$}
    
    \Comment {allocate aggregate intensity $\kappa\in\{0,1\}$ to a set of available players $\mathcal{I}$}
    
    \If{$\kappa = 0$} \Comment {let all available players free-ride}
        \ForAll{$n\in \mathcal{I}$}
            \State $\left. k_n \right\vert_{[a_L,a_R)}\gets 0$
            \State $\left. u_n \right\vert_{[a_L,a_R)}\gets \bar{u}$
        \EndFor
    \ElsIf{$\kappa = 1 = \abs{\mathcal{I}} \eqqcolon \abs{\{n\}}$} \Comment {let the only available player explore}
            \State $\left. k_n \right\vert_{[a_L,a_R)}\gets 1$
            \State $\left. u_n \right\vert_{[a_L,a_R)}\gets \bar{u}$
    \Else \Comment {let one of the $\abs{\mathcal{I}}\geq 2$ available player explore}
        \State $(a_{M-}, a_{M+}, \check{u})\gets \Call{Split}{\abs{\mathcal{I}}, a_L,a_R,\bar{u}}$
        \State $n\gets\min{\mathcal{I}}$
        \State $\left. k_n \right\vert_{[a_L,a_{M-})\cup[a_{M+},a_R)}\gets 1$
        \State $\left. k_n \right\vert_{[a_{M-},a_{M+})}\gets 0$
        \State $\left. u_n \right\vert_{[a_L,a_R)}\gets \check{u}$
        \State $\left. \bar{u}_{\neg n} \right\vert_{[a_L,a_R)}\gets \frac{\abs{\mathcal{I}}\bar{u} - u_n}{\abs{\mathcal{I}} - 1}$ \label{algline:avg-others}
        \Comment {the average payoff among the rest of the players}
    
        \Statex
    
        \State \Call{ActionAssignment}{$\mathcal{I}\setminus\{n\}$, $\kappa$, $a_{M-}$, $a_{M+}$, $\bar{u}_{\neg n}$}
        \State \Call{ActionAssignment}{$\mathcal{I}\setminus\{n\}$, $\kappa - 1$, $a_L$, $a_{M-}$, $\bar{u}_{\neg n}$}
        \State \Call{ActionAssignment}{$\mathcal{I}\setminus\{n\}$, $\kappa - 1$, $a_{M+}$, $a_R$, $\bar{u}_{\neg n}$}
    
    \EndIf
    \EndProcedure
    \end{algorithmic}
\end{algorithm}

\subsubsection*{Step 3: Ensuring mutually best responses.}

Because each \(u_n\) is once differentiable on \(\R_{++}\),
twice differentiable except for the switch points,
and satisfies the normal reflection condition~\eqref{eq:normal-reflection},
our characterization of MPE in Lemma~\ref{lem:char-mpe} states that
\((k_1,\ldots, k_N)\) constitutes an MPE
with \(u_n\) being the associated payoff function of player \(n\)
if \(u_n\) solves the HJB equation
\[
    u_n(a) = 1 + K_{\neg n}(a)\beta(a, u_n) + \max_{k_n\in[0,1]}k_n\{\beta(a, u_n) - 1\}
\]
for each continuity point of \(u_n''\),
and the constructed strategy maximizes the right-hand side.

In other words,
we need to verify for all \(a > 0\),
\(k_n(a) = 1\) if \(\beta(a, u_n) > 1\),
and \(k_n(a) = 0\) if \(\beta(a, u_n) < 1\).
Then, from our construction of \(u_n\),
we can conclude that the HJB equation is satisfied.

On \((a^\flat_N, +\infty)\),
\(u_n(a) = 1\) gives \(\beta(a, u_n) = 0 < 1\),
therefore \(k_n(a) = 0 \) is optimal.
On \([0, a^\sharp_N)\),
the argument is exactly the same as in the symmetric MPE.
Lastly, on \((a^\sharp_N, a^\flat_N)\),
we have \(K = 1\) by construction.
The monotonicity of \(\bar{u}\)
and the construction of \(u_n\)
implies \(1 \leq u_n(a) < \bar{u}(a^\sharp_N) = 2 - 1/N\) for each player \(n\).
Our construction of \(u_n\)
ensures \(k_n(a) = 1\) if and only if \(u_n(a) = \beta(a,u)\),
and \(k_n(a) = 0\) if and only if \(u_n(a) = 1 + \beta(a,u)\)
for each \(x\in(a^\sharp_N, a^\flat_N)\).
Suppose \(\beta(a, u_n) > 1\),
which implies \(1 + \beta(a, u_n) > 2 > u_n(a)\).
Then by construction it must be \(u_n(a) = \beta(a,u)\),
and hence the constructed strategy \(k_n(a) = 1\) is optimal.
On the contrary, if \(\beta(a,u_n) < 1\),
which implies \(\beta(a, u_n) < u_n(a)\),
then by construction it must be \(u_n(a) = 1 + \beta(a, u_n)\),
and hence the constructed strategy \(k_n(a) = 0\) is optimal.

\subsubsection*{Comparison with symmetric MPE.}

Note that \(\bar{u}\) solves ODE
\[
    u'' = \max\{u/N, \min\{1, u - (1 - 1/N)\}\}/\rho + \theta u'
\]
on \((0, a^\flat_N)\),
while the average payoff function \(U^\dagger_N\) of symmetric MPE solves
\[
    u'' = \max\{u/N, 1\}/\rho + \theta u'
\]
on \((0, \tilde{a}_N)\),
with value matching and smooth pasting
at \(a^\flat_N\) and \(\tilde{a}_N\), respectively.
Obviously the right-hand side in the first equation
is weakly smaller.
We can then verify Condition~\ref{item:lem-welfare-comparison-4}
in Lemma~\ref{lem:welfare-comparison}
with \(\breve{u} = \bar{u}(a^\sharp_N) = \min\{\bar{u}(0), 2 - 1/N\}\)
and conclude that
\(\tilde{a}_N < a^\flat_N\)
and \(\bar{u} > U^\dagger_N\) on \([0, a^\flat_N)\).
Similarly, we can also compare the first equation
with the ODE in the cooperative problem \(u'' = u/(N\rho) + \theta u'\)
and conclude \(a^\flat_N < a^*_N\).
\qed

\begin{lemma}[\textsc{Split}]
    Given a strictly decreasing function \(\bar{u}:[a_L, a_R]\to [u_L, u_R]\),
    with \(\bar{u}(a_L) = u_L\) and \(\bar{u}(a_R) = u_R \geq 1\),
    which satisfies the average payoff ODE
    \(u = f + \rho(u '' - \theta u')\)
    on \((a_L, a_R)\)
    with \(0 < f < 1\) and \(\rho > 0\),
    there exist \(a_L < a_{M-} < a_{M+} < a_R\),
    and a function \(\check{u}:[a_L, a_R]\to [u_L, u_R]\) 
    continuously differentiable on \((a_L,a_R)\)
    such that
    \(\check{u}(a_L) = \bar{u}(a_L)\);
    \(\check{u}'(a_L) = \bar{u}'(a_L)\);
    \(\check{u}(a_R) = \bar{u}(a_R)\);
    \(\check{u}'(a_R) = \bar{u}'(a_R)\);
    \(\check{u}\) solves the explorer ODE \(u = \rho (u '' - \theta u')\) on \((a_L, a_{M-})\cup (a_{M+}, a_R)\)
    and solves the free-rider ODE \(u = 1 + \rho (u '' - \theta u')\) on \((a_{M-}, a_{M+})\). \label{lem:split}
    \end{lemma}
    
\begin{proof}
    For any \(a_L \leq a_1 \leq a_2 \leq a_R\),
    consider \(u(\, \cdot \mid a_1,a_2 \,)\) of class \(C^1((a_L, a_R))\)
    that solves the explorer ODE \(u = \rho (u'' - \theta u')\) on \((a_L, a_1)\cup (a_2, a_R)\),
    and the free-rider ODE \(u = 1 + \rho (u'' - \theta u')\) on \((a_1, a_2)\),
    with the initial conditions
    \(u(a_R) = \bar{u}(a_R)\)
    and \(u'(a_R) = \bar{u}'(a_R)\).
    The existence and uniqueness of \(u(\, \cdot \mid a_1,a_2 \,)\) is guaranteed by standard results.
    We write \(u_L^0(a_1, a_2) \coloneqq u(\, a_L \mid a_1,a_2 \,),\)
    and \(u_L^1(a_1, a_2) \coloneqq u'(\, a_L \mid a_1,a_2 \,)\)
    for the value and the derivative of  \(u(\, \cdot \mid a_1,a_2 \,)\) evaluated at \(a_L\).

    Note that
    the functions \(u_L^0, u_L^1:\{\,(a_1, a_2) \mid a_L \leq a_1 \leq a_2 \leq a_R\,\}\to\R_{+}\) are continuous (see, e.g.,\ Theorem 2.14 in \textcite{Barbu:2016}).
    Moreover, 
    from Lemma~\ref{lem:ode-ranking-right} below
    we know that they have the following properties:
    \(u_L^0(a, a) > u_L\),
    \(u_L^1(a, a) < \bar{u}'(a_L) < 0\) for all \(a\in[a_L, a_R]\);
    \(u_L^0(a_L, a_R) < u_L\),
    \(0 \geq u_L^1(a_L, a_R) > \bar{u}'(a_L)\);
    \(u_L^0(a_1, a_2)\) is strictly increasing in \(a_1\)
    and strictly decreasing in \(a_2\).

    By construction, \(u(\, \cdot \mid a_1,a_2 \,)\) and \(\bar{u}\)
    match value and derivative at \(a_R\).
    Next we choose \(a_1\) and \(a_2\) so that 
    they match value and derivatives at \(a_L\) as well.

    Because \(u_L^0(a_L, a_R) < u_L\) and \(u_L^0(a_R, a_R) > u_L\),
    there exists a unique \(\hat{a}_1\in(a_L, a_R)\)
    such that \(u_L^0(\hat{a}_1, a_R) = u_L\),
    and \(u_L^0(a_1, a_R) < u_L\) for all \(a_1 \in [a_L,\hat{a}_1)\).
    Therefore, for each \(a_1\in[a_L, \hat{a}_1]\),
    because \(u_L^0(a_1, a_1) > u_L\),
    there exists a unique \(\check{a}_2(a_1)\in(a_1, a_R]\),
    such that \(u_L^0(a_1, \check{a}_2(a_1)) = u_L\),
    with \(\check{a}_2(\hat{a}_1) = a_R\) by the definition of \(\hat{a}_1\).
    It is obvious that the function \(\check{a}_2\) is continuous
    on its domain \([a_L, \hat{a}_1]\).

    In other words,
    \(u(\, \cdot \mid \hat{a}_1, a_R \,)\) solves
    the free-rider ODE on \((\hat{a}_1, a_R)\)
    and the explorer ODE on \((a_L, \hat{a}_1)\),
    while \(u(\, \cdot \mid a_L, \check{a}_2(a_L) \,)\) solves
    the explorer ODE on \((\check{a}_2(a_L),a_R)\)
    and the free-rider ODE on \((a_L, \check{a}_2(a_L))\),
    with both \(u(\, \cdot \mid \hat{a}_1, a_R \,)\)
    and \(u(\, \cdot \mid a_L, \check{a}_2(a_L) \,)\)
    having the same value as \(\bar{u}\) at \(a_L\) and \(a_R\),
    and the same derivatives as \(\bar{u}\) at \(a_R\).

    From Lemma~\ref{lem:ode-ranking-right} below
    we can conclude
    that \(u_L^1(\hat{a}_1, a_R) < \bar{u}'(a_L) < u_L^1(a_L, \check{a}_2(a_L))\).\footnote{
        Write \(\tilde{u}(\cdot)\coloneqq u(\, \cdot  \mid  \hat{a}_1, a_R \,)\).
        First, we show
        \(u_L^1(\hat{a}_1, a_R) \coloneqq \tilde{u}'(a_L) \leq \bar{u}'(a_L)\).
        Since \(\tilde{u}\) solves the free-rider ODE on \((\hat{a}_1, a_R)\),
        with the same initial conditions at \(a_R\) as \(\bar{u}\),
        from Lemma~\ref{lem:ode-ranking-right} we know that
        \(\tilde{u}(\hat{a}_1) < \bar{u}(\hat{a}_1)\).
        Then we must have 
        \(\tilde{u}(a) < \bar{u}(a)\) for all \(a\in(a_L, \hat{a}_1)\)
        for the following reason.
        Suppose by contradiction this is not the case,
        then there exists some \(\breve{a} \in(a_L, \hat{a}_1)\) such that
        \(\tilde{u}(\breve{a}) = \bar{u}(\breve{a})\)
        and \(\tilde{u}'(\breve{a}) \leq \bar{u}'(\breve{a}) \leq 0\).
        Now, since \(\tilde{u}\) solves
        the explorer ODE on \((a_L, \breve{a})\),
        Lemma~\ref{lem:ode-ranking-right}
        implies that \(\tilde{u}(a_L) > \bar{u}(a_L)\),
        which contradicts the fact that
        \(\tilde{u}(a_L) = \bar{u}(a_L)\),
        guaranteed by our choice of \(\hat{a}_1\).
        Therefore we have 
        \(\tilde{u}(a) < \bar{u}(a)\) for all \(a\in(a_L, \hat{a}_1)\),
        which together with \(\tilde{u}(a_L) = \bar{u}(a_L)\),
        implies \(\tilde{u}'(a_L) \leq \bar{u}'(a_L)\).

        Next, we show that
        this inequality is strict.
        Suppose by contradiction that \(\tilde{u}'(a_L) = \bar{u}'(a_L)\).
        Then \(u\) and \(\bar{u}\)
        have the same values and derivatives at \(a_L\).
        From the ODEs which \(\tilde{u}\) and \(\bar{u}\) satisfy
        on \((a_L, \hat{a}_1)\),
        we can conclude \(\tilde{u}''(a_L) > \bar{u}''(a_L)\).
        This contradicts the fact that \(\tilde{u}(a) < \bar{u}(a)\)
        on \((a_L, \hat{a}_1)\),
        and therefore we must have
        \(\tilde{u}'(a_L) < \bar{u}'(a_L)\).
        The other inequality can be derived analogously.
        }
    Therefore, there exists some \(\check{a}_1\in(a_L, \hat{a}_1)\) such that
    \(u_L^1(\check{a}_1, \check{a}_2(\check{a}_1)) = \bar{u}'(a_L)\) from the continuity of \(u_L^1\) and \(\check{a}_2\).
    Because \(\check{a}_1 < \hat{a}_1\),
    we have \(u_L^0(\check{a}_1, a_R) < u_L\),
    which implies \(\check{a}_2(\check{a}_1) \neq a_R\)
    and thus it must be \(\check{a}_2(\check{a}_1) < a_R\).

    Finally, let
    \(a_{M-}\coloneqq \check{a}_1\),
    \(a_{M+}\coloneqq \check{a}_2(\check{a}_1)\)
    and
    \(\check{u}(\cdot) \coloneqq u(\, \cdot \mid a_{M-},a_{M+} \,)\).
    We have shown that \(\check{u}(a_L) = u_L\),
    \(\check{u}'(a_L) = \bar{u}'(a_L)\),
    and by construction \(a_L < a_{M-} < a_{M+} < a_R\).
    The range of \(\check{u}\) follows from the monotonicity of \(\check{u}\),
    which is implied by Lemma~\ref{lem:ode-ranking-right}
    with \(\check{u}'(a_R) = \bar{u}'(a_R) \leq 0\).
\end{proof}

\begin{lemma}
    Let \(u(\, \cdot \mid f, u^0, u^1 \,)\) be the solution of
    the ODE \(u = f + \rho (u'' - \theta u')\)
    with the initial conditions \(u(a_R) = u^0\) and \(u'(a_R) = u^1\),
    for some \(\rho > 0\) and \(f,\theta\in\R\).
    For any \(a_L < a_R\),
    \(u(\, a_L \mid f, u^0, u^1 \,)\)
    is strictly increasing in \(u^0\)
    and strictly decreasing in \(f\) and \(u^1\),
    whereas \(u'(\, a_L \mid f, u^0, u^1 \,)\)
    is strictly decreasing in \(u^0\)
    and strictly increasing in \(f\) and \(u^1\).
    Moreover, if \(u^0 \geq f\) and \(u^1 \leq 0\),
    then \(u'(\, a_L \mid f, u^0, u^1 \,) \leq 0\),
    with \(u'(\, a_L \mid f, u^0, u^1 \,) = 0\)
    if and only if
    \(u^0 = f\) and \(u^1 = 0\).
    \label{lem:ode-ranking-right}
\end{lemma}
\begin{proof}
    Let \(\eta\) denote either \(f\), \(-u^0\), or \(u^1\).
    For \(a_L < a_R\),
    to show \(u'(\, a_L \mid \eta \,)\) is strictly increasing in \(\eta\),
    let \(\eta_1 < \eta_2\).
    For the purpose of contradiction,
    suppose \(u'(\, a_L \mid \eta_1 \,) \geq u'(\, a_L \mid \eta_2 \,)\).
    Let \(\hat{a} = \sup\{\,a < a_R \mid u'(\, a \mid \eta_1 \,) = u'(\, a \mid \eta_2 \,)\,\}\).

    For \(\eta\) being either \(f\) or \(-u^0\),
    because \(\eta_1 < \eta_2\)
    and \(u''(\, a_R \mid \eta \,) = \frac{u^0 - f}{\rho} + \theta u^1\),
    we have \(u''(\, a_R \mid \eta_1 \,) > u''(\, a_R \mid \eta_2 \,)\).
    Therefore, we have \(u'(\, a \mid \eta_1 \,) < u'(\, a \mid \eta_2 \,)\)
    for all \(a\in(a_R - \epsilon, a_R)\) for some \(\epsilon > 0\),
    which obviously also holds 
    when \(\eta\) denotes \(u^1\)
    by the continuity of \(u'(\, \cdot \mid \eta \,)\).
    Therefore,
    we have \(a_L \leq \hat{a} < a_R\),
    and from the continuity of \(u'(\, \cdot \mid \eta \,)\),
    we have \(u'(\, \hat{a} \mid \eta_1 \,) = u'(\, \hat{a} \mid \eta_2 \,)\).

    Because \(u'(\, a \mid \eta_1 \,) < u'(\, a \mid \eta_2 \,)\)
    for all \(a\in(\hat{a}, a_R)\),
    we have
    \begin{align*}
        u(\, \hat{a} \mid \eta_1 \,)
        &= u(\, a_R \mid \eta_1 \,) - \int_{\hat{a}}^{a_R} u'(\, a \mid \eta_1 \,) \dd{a} \\
        &> u(\, a_R \mid \eta_2 \,) - \int_{\hat{a}}^{a_R} u'(\, a \mid \eta_2 \,) \dd{a}
        = u(\, \hat{a} \mid \eta_2 \,),
    \end{align*}
    which implies
    \begin{align*}
        u''(\, \hat{a} \mid \eta_1 \,)
        &= \frac{u(\, \hat{a} \mid \eta_1 \,) - \left.f\right\vert\eta_1}{\rho} + \theta u'(\, \hat{a} \mid \eta_1 \,) \\
        &> \frac{u(\, \hat{a} \mid \eta_2 \,) - \left.f\right\vert\eta_2}{\rho} + \theta u'(\, \hat{a} \mid \eta_2 \,)
        =u''(\, \hat{a} \mid \eta_2 \,).
    \end{align*}
    But this is a contradiction,
    as \(u'(\, \hat{a} \mid \eta_1 \,) = u'(\, \hat{a} \mid \eta_2 \,)\)
    together with \(u'(\, a \mid \eta_1 \,) < u'(\, a \mid \eta_2 \,)\) for all \(a\in(\hat{a}, a_R)\)
    implies
    \(u''(\, \hat{a} \mid \eta_1 \,) \leq u''(\, \hat{a} \mid \eta_2 \,)\).

    As \(a_L\) is chosen arbitrarily,
    we have shown that for any \(a < a_R\),
    \(u'(\, a \mid \eta \,)\) is strictly increasing in \(\eta\).
    Therefore,
    as \(u(\, a_L \mid \eta \,) = u(\, a_R \mid \eta \,) - \int_{a_L}^{a_R} u'(\, a \mid \eta \,) \dd{a}\),
    we have \(u(\, a_L \mid \eta \,)\) is strictly decreasing in \(\eta\).

    Finally, if \(u^0 = f\) and \(u^1 = 0\),
    then \(u(\, a \mid f, u^0, u^1 \,) = f\) is the solution of the ODE
    and hence \(u'(\, a_L \mid f, u^0, u^1 \,) = 0\) for any \(a_L \leq a_R\).
    Otherwise,
    if either \(u^0 > f\) or \(u^1 < 0\),
    from the strict monotonicity of \(u'(\, a_L \mid f, u^0, u^1 \,)\)
    in \(u^0\) and \(u^1\),
    we have \(u'(\, a_L \mid f, u^0, u^1 \,) < 0\).
\end{proof}

\subsection{Proof of Proposition~\ref{prop:bound-two-player-mpe}}
\begin{proof}
    The average payoff \(\hat{u}\)
    in any MPE must be once continuously differentiable and
    satisfy
    the ODE
    \(u'' = g(u, u')\) for some \(g\)
    respecting Feynman-Kac equation~\eqref{eq:feynman-kac-x}
    whenever \(\hat{u} > 1\) and \(\hat{u}''\) is continuous.
    For \(N=2\),
    the characterization of best responses gives
    the following minimal requirements on \(g\).
    For each \(1 < u < 2\),
    \(g(u, u')\) can only be either
    \(\frac{u-1/2}{\rho} + \theta u'\) (one player explores and the other free-rides)
    or \(\frac{1}{\rho} + \theta u'\) (both players choose a common interior allocation);
    for each \(u > 2\),
    \(g\) can only be either
    \(\frac{u-1/2}{\rho} + \theta u'\)  (one player explores and the other free-rides)
    or \(\frac{u}{2\rho} + \theta u'\) (both players explore).
    The asymmetric MPE we construct in Proposition~\ref{prop:asymmetric-mpe}
    adopts \(g(u, u') = \max\{u/2, \min\{1, u-1/2\}\}/\rho + \theta u'\),
    which is the minimal \(g\)
    that satisfies these constraints,
    and therefore achieves the highest payoff among all MPE by Lemma~\ref{lem:welfare-comparison}.
\end{proof}

\subsection{Proof of Proposition~\ref{prop:pareto}}
\begin{proof}
    Choose a partition \(\{a_j\}_{j=1}^{m+1}\) of \([a^\sharp_N, a^\flat_N]\)
    in the construction of the equilibria in Proposition~\ref{prop:asymmetric-mpe},
    so that
    \[
        \max_{1\leq j \leq m}\abs{\bar{u}(a_{j+1}) - \bar{u}(a_j)} \leq \delta
        \coloneqq \frac{1}{2}\min_{a^\sharp_N \leq a \leq \tilde{a}_N}\{\bar{u}(a) - U^\dagger_N(a)\},
    \]
    and \(\abs{a_m - a_{m+1}} < \epsilon\).
    Because the average payoff \(\bar{u}\)
    and the players' payoff functions
    are monotone and coincide at the endpoints of the subintervals \([a_j, a_{j+1}]\),
    we have \(\abs{u_n(a) - \bar{u}(a)} \leq \delta\) for \(a\in [a^\sharp_N, \tilde{a}_N]\)
    for all player \(n\).
    Therefore,
    we have
    \(u_n\geq \bar{u} - \delta > U^\dagger_N\) on \([a^\sharp_N, \tilde{a}_N)\).
    Also,
    as \(u_n(a_m) = \bar{u}(a_m) > 1\) for all player \(n\),
    we have
    \(u_n > 1 = U^\dagger_N\) on 
    \([\tilde{a}_N, a_m]\supset[\tilde{a}_N, a^\flat_N - \epsilon]\).
    Lastly, \(u_n = \bar{u} > U^\dagger_N\) on \([0, a^\sharp_N)\),
    if it is not empty.
\end{proof}

\printbibliography

\end{document}